\documentclass[preprint,12pt,authoryear]{elsarticle}
\pdfoutput=1

\usepackage{subcaption}
\usepackage{caption}
\usepackage{seqsplit}
\usepackage{url}

\usepackage{xspace}
\usepackage{tikz}
\usetikzlibrary{shapes.geometric, arrows.meta, positioning, fit, backgrounds, calc}

\usepackage{amsmath}

\newcommand\gridfire{\texttt{GridFire}\xspace}
\newcommand\gridfires{\texttt{GridFire's}\xspace}
\newcommand\pynucastro{\texttt{pynucastro}\xspace}

\newcommand\qsev{\texttt{MultiscalePartitioningEngineView}\xspace}

\newcommand\graphengine{\texttt{GraphEngine}\xspace}
\newcommand\dsep{\texttt{DSEP}\xspace}
\newcommand\mesa{\texttt{MESA}\xspace}
\newcommand\mesas{\texttt{MESA's}\xspace}
\newcommand\cvode{\texttt{CVODE}\xspace}
\newcommand\kinsol{\texttt{KINSOL}\xspace}

\journal{Astronomy and Computing}

\begin{document}

\begin{frontmatter}
\title{\gridfire: A new, open source, nuclear network for stellar structure and evolution}
\author[1]{Emily M. Boudreaux\corref{cor1}}
\ead{Emily.Boudreaux@dartmouth.edu}
\author[1]{Aaron Dotter}
\ead{Aaron.Dotter@dartmouth.edu}

\affiliation[1]{organization={Department of Physics and Astronomy}, addressline= 6127 Wilder Laboratory, Dartmouth College, postcode={03755}, city={Hanover}, country={USA}}

\begin{abstract}
  We present a novel nuclear network implementation, \gridfire, which has been
  developed with a focus on efficiency, ease of use, and automatic physical
  extension. In this first paper we provide a detailed overview of \gridfires
  numerics, performance characteristics, and a comparison to pre-existing
  nuclear network implementations \pynucastro and \mesas \texttt{net} module. We
  find that generally \gridfire performs similarly to and requires significantly
  less user input and setup than these current generation nuclear
  network codes. Further, \gridfire has been developed with a flexible
  ``view'' based physics engine that allowing researchers to implement domain
  specific physics without modifying the underlying source code. \gridfire has
  been released under an open-source license, GPL v3.0, and will eventually be
  incorporated into the forthcoming 4D-STAR stellar structure evolution code.
\end{abstract}
\end{frontmatter}

\section{Introduction}\label{sec:intro}
Stellar structure and evolutionary (SSE) models underpin a large section
of our current astrophysical model. Traditionally, these models operate in two
phases: one, structure equations, and two, microphysics solutions. Solutions to
the structure equations serve to constrain the profiles of thermodynamic
variables throughout the domain. However, these solutions rely on input
microphysics. Some of this microphysics, most notably radiative and conductive
opacities, is generally pretabulated, while other parts, such as the rate of
energy generation and chemical evolution due to nuclear burning, tend to be
calculated, at least partially, on-the-fly. Efficient and robust nuclear
networks are therefore essential in order to evolve the chemical
abundance of species, track the energy input into the model, and track energy
lost to free-streaming neutrinos \citep{Halprin1985, Paxton2011, Paxton2019,
Bowman2023,Farag2024}. These networks can take on two primary forms: one, static
\citep[e.g. the approx networks in \mesa and
\texttt{pynucastro},][]{Timmes1999,Paxton2011, SmithClark2022}, and two, dynamic.
In a static network a fixed set of equations describing the relation between
abundance, temperature, and density are solved over some time step.  Static
networks have the advantage of often being both more performant and easier to
implement. However, this network paradigm may struggle to evolve through
significant model phase changes \citep{Timmes1999, Mocak2009, Mocak2010}.
Further, as their structure cannot alter in response to model state, static
networks may be unable to leverage optimizations which are only valid during
certain stages of model evolution \citep[i.e. as temperature evolves from energy
input into the system, e.g.][]{Mocak2009, Jiang2021}. Dynamic nuclear networks
seek to address these shortcomings while maintaining as much performance as
possible.

Dynamic nuclear networks, as a category, implement some strategy to adapt their
topology during model evolution. This results in networks which can detect both
cycles and low flow edges. Once these detections are made dynamic networks may
apply optimizations such as equilibrium approximations for cycles and pruning of
low flow edges. Further, these structural choices need not be fixed over the
life of a model; rather, model state can be probed to inform choices about when
network structure should be updated. This more freeform handling of nuclear
burning\footnote{We use the term burning colloquially to refer to nuclear fusion
processes throughout this article.} does tend to increase the per-time step cost;
however, at the advantage of a potentially much more well conditioned system,
and a system which can more easily adapt to dramatic phase changes during model
evolution. 

\gridfire is a new dynamic nuclear network implementation with the specific
design goal of modeling main sequence nuclear burning with minimum user input.
A key design goal is to maintain both a physically robust tool while keeping the
barrier to entry as low as possible. The architecture of \gridfire can roughly
be broken down into two segments: engines, responsible for local microphysics,
and solvers, responsible for stepping engines through time. Engines and solvers
are discussed in more detail in sections \ref{sec:engines} and \ref{sec:solvers}
respectively; however, in brief engines are automatically constructed
directional hypergraphs representing some network topology. Further, \gridfire
defines the concept of an ``Engine View'' (views) which can modify the exposed
state of an engine (\S \ref{sec:engines:views}). Currently we include a view
which can automatically detect quasi-static equilibrium clusters (\S
\ref{sec:views:qse}). Solvers on the other hand accept an engine as a
black-box and may implement single- or multi-zone evolution prescriptions based
on the physics exposed by an engine. \gridfire is packaged with a single zone
solver as well as a simplistic multi-zone solver. The bundled engines, views,
and solvers allow \gridfire to reliably explore the entire main-sequence
burning regime robustly and efficiently from only a limited subset of seed
species.

\begin{figure}
  \centering
  \resizebox{\linewidth}{!}{\begin{tikzpicture}[
    node distance=0.8cm and 1.0cm,
    font=\sffamily\footnotesize,
    >=Latex,
    startstop/.style={
        rectangle, rounded corners=3mm, minimum width=3cm, minimum height=0.8cm,
        text centered, draw=black!80, fill=red!10, thick
    },
    process/.style={
        rectangle, minimum width=3.5cm, minimum height=0.8cm,
        text centered, draw=black!80, fill=blue!5, thick
    },
    subroutine/.style={
        rectangle, minimum width=3.5cm, minimum height=0.8cm,
        inner xsep=4mm, 
        text centered, draw=black!80, fill=blue!15, thick,
        path picture={
            \draw[black] ([xshift=2mm]path picture bounding box.north west) -- ([xshift=2mm]path picture bounding box.south west);
            \draw[black] ([xshift=-2mm]path picture bounding box.north east) -- ([xshift=-2mm]path picture bounding box.south east);
        }
    },
    decision/.style={
        diamond, 
        aspect=1.8, 
        minimum width=1.5cm, 
        minimum height=0.8cm,
        text centered, draw=black!80, fill=green!10, inner sep=1pt
    },
    io/.style={
        trapezium, 
        trapezium left angle=80, trapezium right angle=100, 
        minimum width=0cm, 
        minimum height=0.8cm,
        text centered, draw=black!80, fill=orange!10,
        inner xsep=4mm 
    },
    container/.style={
        draw=gray!50, dashed, rounded corners, inner sep=0.5cm, fill=white, fill opacity=0.3
    },
    line/.style={draw, ->, thick, black!80},
    connector/.style={draw, ->, thick, black!80, rounded corners},
    label_text/.style={font=\scriptsize\itshape, text=gray!80, anchor=west}
]
\node (define_comp) [startstop] {\textbf{Define Composition}};
\node (init_conditions) [startstop, left=1cm of define_comp] {\textbf{Set initial conditions (T, $\rho$)}};
\node (build_netin) at ($(define_comp)!0.75!(init_conditions) + (0, -1.5)$)[process] {Build \texttt{NetIn}(s)};
\node (select_policy) [process, below=2cm of define_comp] {Select Policy};
\node (construct_policy) [subroutine, below=1cm of select_policy] {Call \texttt{policy.construct()}};

\node (engine) at ($(construct_policy) + (+2, -1.5)$) [io] {Engine};
\node (context) at ($(construct_policy) + (-2, -1.5)$) [io] {Context};

\node (solver) [process, below=2.5cm of construct_policy] {Select Solver};
\node (solver_ctx) [process, left=1cm of solver] {Select Solver Context};
\node (solver_options) [process, below=1cm of solver_ctx] {Set Solver Options};

\node (evaluate) at ($(solver)!0.5!(solver_options) + (0, -2.5)$) [subroutine] {Call \texttt{solver.evaluate(ctx, netIn(s))}};

\node (results) [startstop, below=1cm of evaluate] {Results};

\begin{scope}[on background layer]
    \node (construction_results) [container, fit=(engine) (context), label={[anchor=north, font=\bfseries\scriptsize, text=gray, yshift=-1cm]east:Construction Results}] {};
\end{scope}

\draw [connector] (define_comp.south) -- +(0, -1.075) -- (build_netin.east);
\draw [connector] (define_comp.south) -- (select_policy.north);
\draw [connector] (init_conditions) -- +(0, -0.75) -- +(1.275, -0.75) -- (build_netin.north);
\draw [connector] (select_policy.south) -- (construct_policy.north);
\draw [connector] (construct_policy.south) -- +(0, -1.05) -- (engine.west);
\draw [connector] (construct_policy.south) -- +(0, -1.05) -- (context.east);

\draw [connector] (engine.south) -- +(0, -0.5) -- +(-2.0, -0.5) -- (solver.north);
\draw [connector] (context.south) -- +(0, -0.5) -- +(-2.5, -0.5) -- (solver_ctx.north);

\draw [connector] (solver.south) -- +(0, -2) -- +(-2.25, -2) -- (evaluate.north);

\draw [connector] (solver_ctx.south) -- (solver_options.north);
\draw [connector] (solver_options.south) -- +(0, -0.215) -- +(2.25, -0.215) -- (evaluate.north);

\draw [connector] (build_netin) -- +(0, -2) -- +(-3, -2) -- +(-3, -9.90) -- (evaluate.west);

\draw [connector] (evaluate.south) -- (results.north);

\end{tikzpicture}}
  \caption{Prototypical Example of \gridfire usage from a user perspective.}
  \label{fig:user_control_flow}
\end{figure}
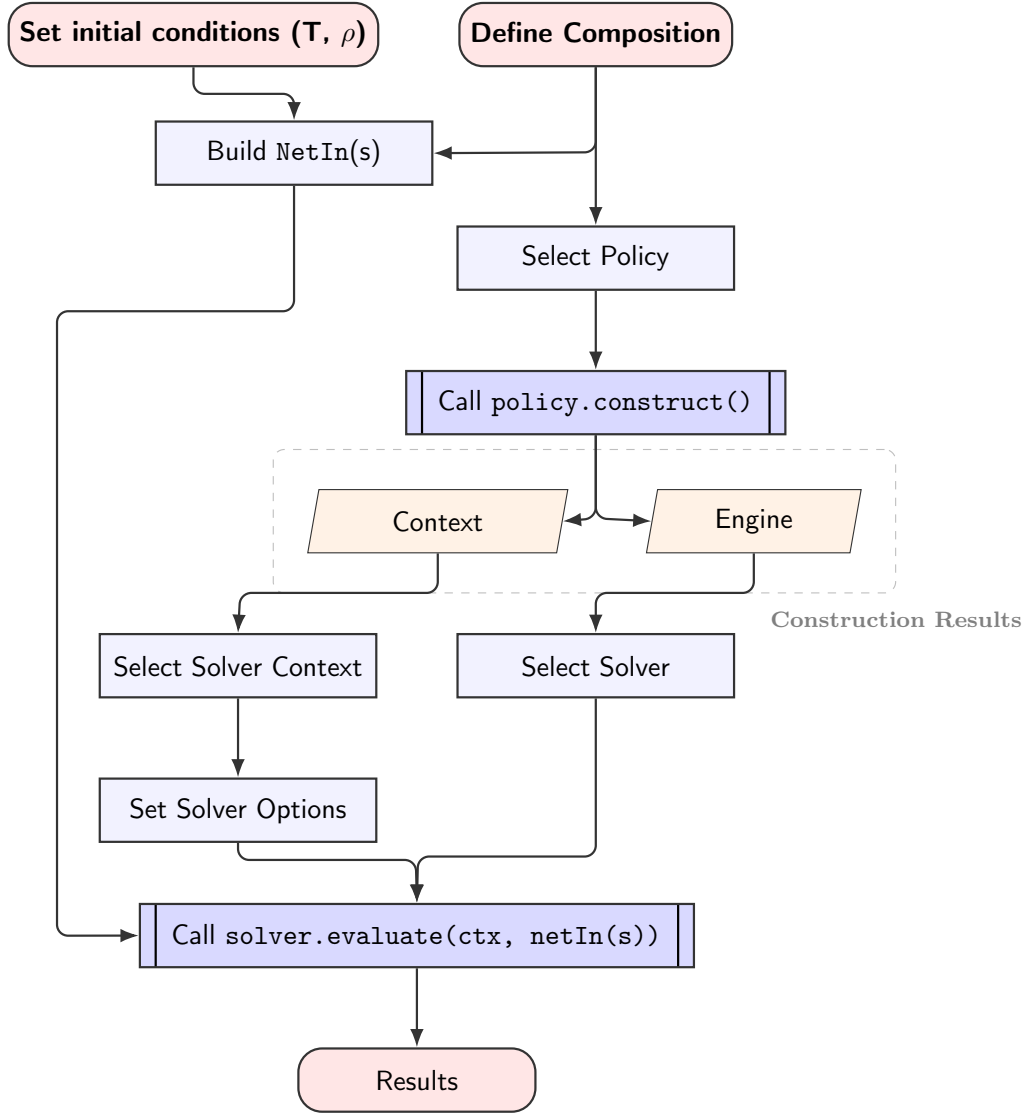

\gridfire is intentionally designed to be easy to use and require minimal user
input (a prototypical example of \gridfire usage may be seen in Figure
\ref{fig:user_control_flow}). Specifically, \gridfire does not require users
define which reactions pathways are turned on or off nor does it require users
to define which species are being tracked. Rather, these topological questions
are generally determined by \gridfire at runtime --- though in cases where a
specific set of reactions are desired that information may be provided
explicitly through the use of \gridfires views system. Further,
\gridfire has been designed from the ground up with extensibility in mind. New
engines, views, and solvers can be easily implemented in either C++ or Python
without modification to the underlying code base. The engine view architecture
in particular enables modification of physics without requiring an overly
detailed knowledge of \gridfires structure or C++. Support for reverse reaction
rates via detailed balance calculations, while currently experimental, is
included. In this paper we will provide a brief overview of how nuclear network
programs operate, detail \gridfires design, its ability to robustly model main
sequence burning, and demonstrate a simplistic big bang nucleosynthesis model
built with \gridfire.

\section{Nuclear Networks}\label{sec:nuclear_network}
Computing an updated abundance profile, specific energy generation, and specific
neutrino losses for a fusing plasma requires knowledge of the thermodynamic
conditions, initial composition, possible reaction pathways, and reaction rates.
Once these are known the process is, theoretically, quite simple; one needs to
solves the ordinary differential equation given in Equation \ref{eqn:burning_1}
as an initial value problem where $\mathbf{f}_{i}(\mathbf{Y}, \rho, T)$ is some
function describing the net molar reaction flow for a species, $i$, at the
current conditions and $N$ is the total number of species.

\begin{equation}\label{eqn:burning_1}
  \begin{aligned}
    \frac{dY_{i}}{dt} &= \mathbf{f}_{i}(\mathbf{Y}, \rho, T) \ \ \forall \ \ i = 1...N \\
    \frac{d\epsilon_{nuc}}{dt} &= \dot{\epsilon}_{nuc}(\mathbf{Y}, \rho, T)
  \end{aligned}
\end{equation}

There are however a number of complicating factors which make this less trivial than
it may seem at face value. For one, knowledge of the molar reaction flow for all
potentially relevant reactions is impractical to compute from first
principles. Rather, reactant count- and density-normalized reaction rates are
generally pulled from pre-tabulated data files. Further, the characteristic
stiffness of these equations, for any physically interesting network, is
pathologically large. This latter issue makes all explicit solvers, and in fact
many implicit solvers struggle.  Stable solutions to these equations therefore
require implicit solvers which are well tuned to extremely stiff systems, such
as backwards difference formula (BDF) family solvers (see \S \ref{sec:solvers} for
more details on the solvers \gridfire implements)

Assuming then that a solver has access to reaction rates for every reaction of
interest Equation \ref{eqn:burning_1} can be expanded into Equation
\ref{eqn:burning_2}, where $c_{i, j}$ is the stoichiometric coefficient for
species $i$ in reaction $j$, $r$ is the total number of reactions tracked,
$N_{A}$ is Avogadro's constant, and $\mathcal{R}_{j}$ is the molar reaction flow
for reaction some reaction $j$.

\begin{equation}\label{eqn:burning_2}
  \begin{aligned}
  \frac{dY_{i}}{dt} &= \sum_{j}^{r}c_{i,j}\mathcal{R}_{j}(\mathbf{Y}, \rho, T) \ \ \forall \ \ i = 1...N \\
  \frac{d\epsilon_{nuc}}{dt} &= -N_{A}c^{2}\sum_{i}^{N}\frac{dY_{i}}{dt}m_{i} 
  \end{aligned}
\end{equation}

\begin{figure}
  \centering
  \includegraphics[width=0.47\textwidth]{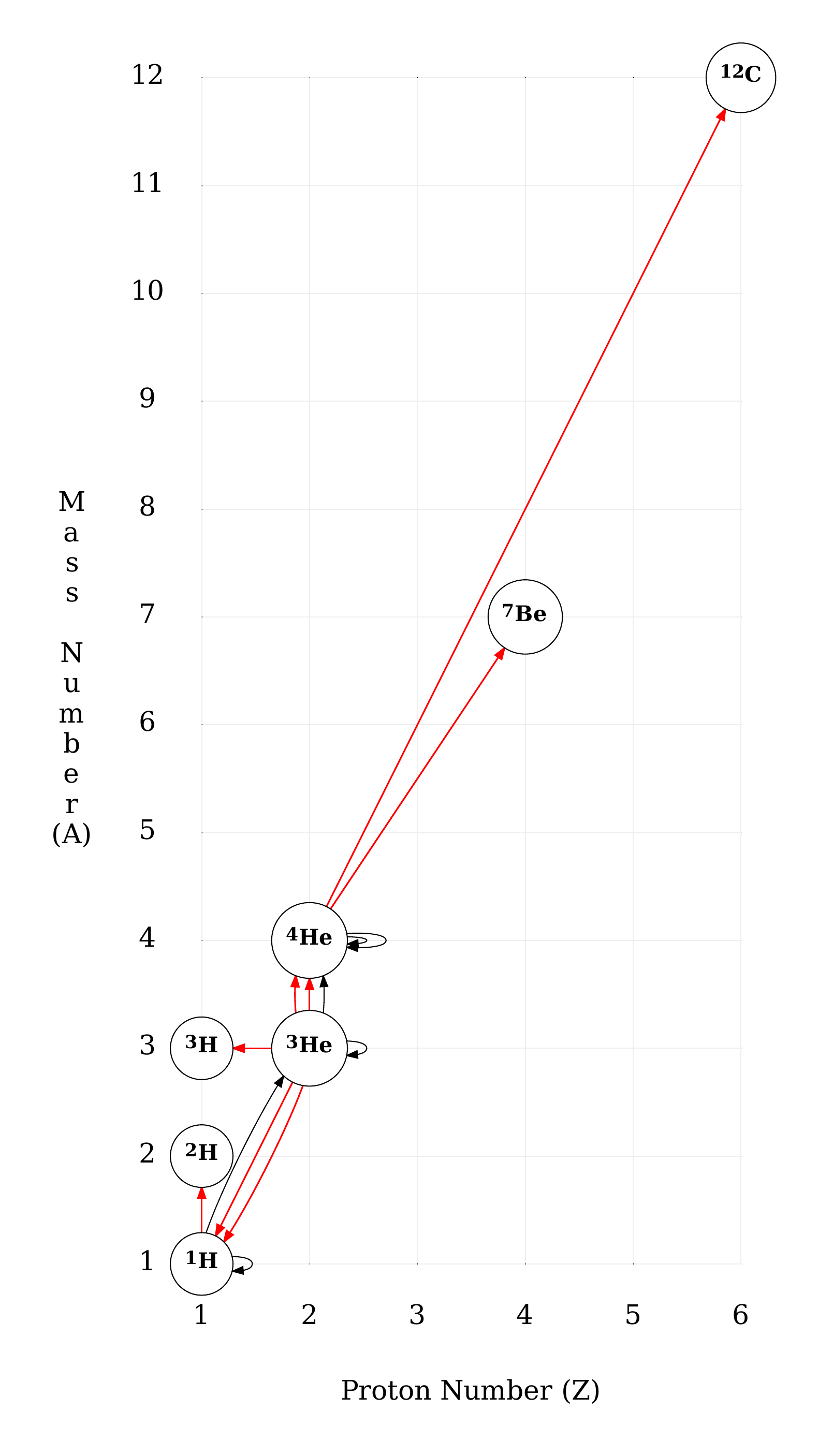}
  \includegraphics[width=0.47\textwidth]{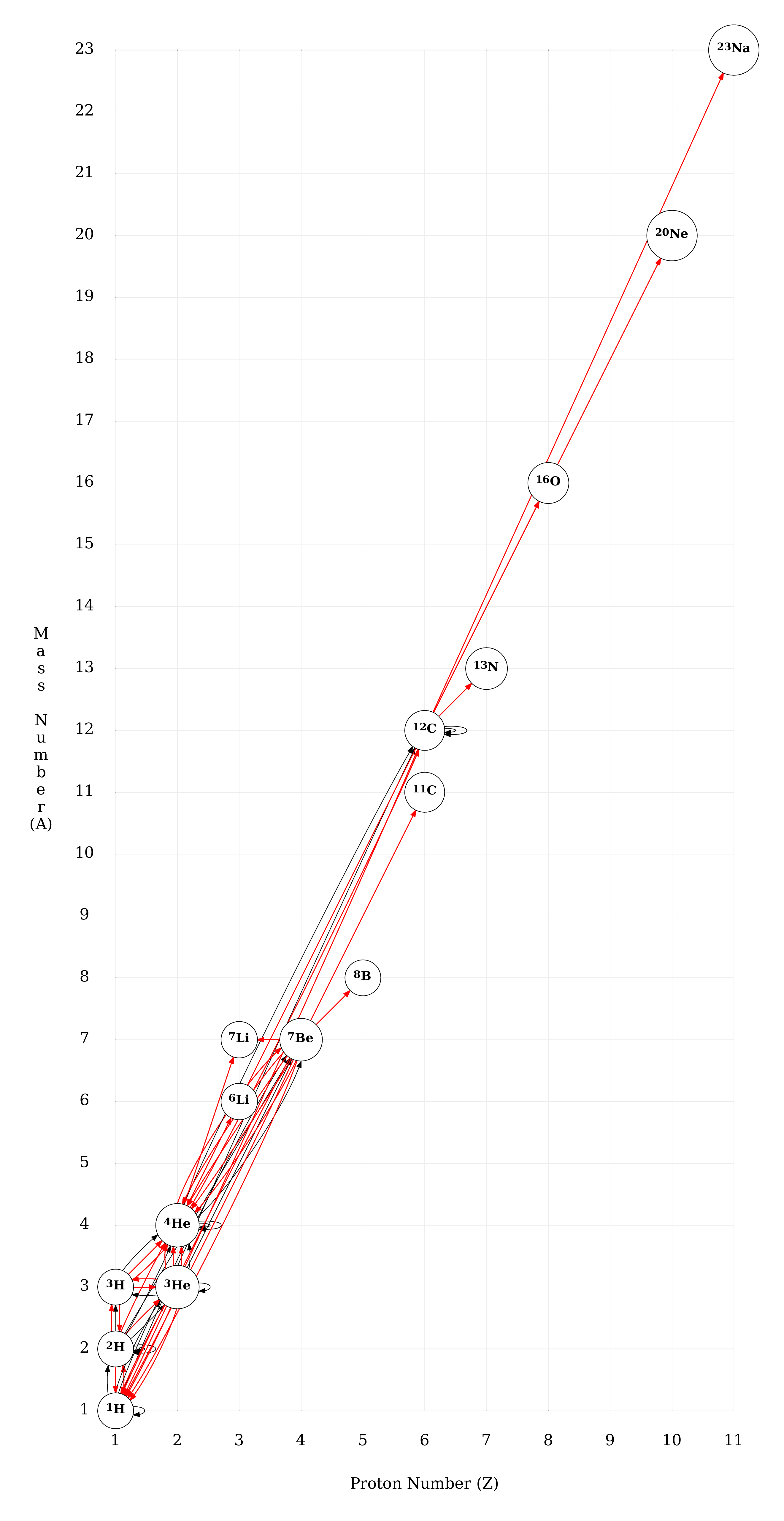}
  \caption{Two simple network topologies automatically generated by
  \gridfire. Black arrows indicate reactant connections whereas red arrows
  indicate products (i.e. for the reaction $p + p \rightarrow d + e^{+}$ there
  would be a self-loop black arrow from H-1 to itself and a red arrow to H-2. For
  clarity neutron-mediated reactions have been omitted.  (left). Single-layer
  network constructed from a seed composition of $^{1}$H, $^{3}$He, and
  $^{4}$He, (right). Two-layer network constructed from a seed composition of
  $^{1}$H, $^{3}$He, and $^{4}$He. Note how with only a single layer deeper of
  construction the topology becomes dramatically more complex.}
  \label{fig:topology}
\end{figure}

While nuclear burning can be modeled as a fixed set of coupled equations, in
order to build a dynamic nuclear burning code it is advantageous to
conceptualize the topology of the network. For some set of reactions and species
a nuclear network is a directional hypergraph (Figure \ref{fig:topology}) where
species form nodes, species sets form hypernodes, and reactions form edges.
Focusing on the topology of the network allows nuclear burning code to perform
optimizations such as low molar reaction flow culling, wherein reactions (edges)
are weighted by their molar reaction flows and, based on some threshold,
reactions whose flow does not exceed this threshold can be removed from the network
along with any species only connected to the greater network through those
reactions. Further, given some at least somewhat extensive underlying database
of reactions, standard graph traversal algorithms may be used to construct a
network topology from some seed composition automatically. Finally, once
constructed, the equations needed to time evolve the composition and energy
generation of a fusing plasma may be constructed directly from the topology.

Time-stepping a network requires a well optimized implicit method
to handle the inherent stiffness of nuclear burning. Further complicating this
picture is a choice that nuclear burning code must make regarding thermodynamic 
stability. While burning, energy is produced which --- physically --- will feed
back into the plasma, potentially heating it and affecting its density.
In non-degenerate matter this will result in an increasing burning
rate over time.  In order to model this effect a prescription for an equation of
state (EOS) is required. Generally nuclear networks take two approaches to this
problem, fully thermodynamically coupled, and operator splitting.
Thermodynamically coupled networks maintain a self-consistent state by ensuring
that temperature and density at every time step properly include the heating
effects of burning \citep[e.g. SkyNet,][]{Lippuner2017}; this comes at the cost of potentially much more expensive
time steps due to both EOS evaluations and additional variable couplings. Conversely,
operator splitting relies on the assumption that there is some bath large enough
to absorb all generated energy over the course of a timestep such that
the temperature and density of the burning media remain fixed \citep[e.g. WinNet,][]{Reichert2023}. Both techniques
are common in astrophysical code bases though non-thermodynamically coupled
networks / those intended to be used in an operator splitting fashion are often
unable to properly evolve through highly dynamic events such as the helium flash
or neutron star mergers.

\section{\gridfires Architecture}\label{sec:design}
\gridfires design makes heavy use of the so called ``strategy pattern''. A
detailed explanation of this pattern was originally provided in
\citet{Gamma1993}; however, in brief this design prioritizes encapsulated
interchangeable components enabling these components to be easily swapped in and
out. Table \ref{tab:strategies} summarizes the major components which \gridfire
implements while Figure \ref{fig:gridfire_flow_chart} demonstrates the
prototypical control flow for a single-zone solver. A major benefit of the
strategy pattern is that, at any stage in this flow, researchers are able to substitute components
more suited to their specific question. For example, we include
\texttt{MultiscalePartitioningEngineView} (for more details on this see \S
\ref{sec:views:qse}), this view implements a modification of the \citet{Hix1999}
quasi static equilibrium prescriptions --- taking groups of species which are
heavily exchanging nucleons and placing them into equilibrium --- If this
behavior is not desirable for some domain that view may simply be removed from
the engine stack. Further details on \gridfires software architecture as well as
usage examples may be found in its
documentation\footnote{https://4d-star.github.io/GridFire/html/index.html}. A
prototypical example of the steps \gridfire takes during an evaluation of a
network over some time and at a fixed temperature and density is given below.

\begin{enumerate}
  \item Collect reactions from the JINA REACLIB database \citep{Cyburt2010}
  \item Starting from some seed composition, perform a breadth first search outward to some, user-specified, maximum depth. Collect all reactions and species.
  \item Use the collected reaction set to construct a \texttt{GraphEngine}
  \item This \texttt{GraphEngine} evaluates all partial derivatives of the entire collected reaction set using auto-differentiation
  \item The sparsity pattern of this Jacobian matrix is computed
  \item Views above \texttt{GraphEngine} then perform tasks. These will be outlined in latter sections.
  \item The entire engine stack is passed to a backwards difference formula (BDF) solver (\cvode) which periodically evaluates rates and partial derivatives.
\end{enumerate}

\begin{table}[htb!]
\centering
\footnotesize
\begin{tabular}{r | l p{0.45\textwidth}}
\hline
\textbf{Name} & \textbf{Namespace} & \textbf{Purpose} \\
\hline
\texttt{DynamicEngine} & \texttt{gridfire::engine} & Abstract engine type; all dynamic engines can generate the right-hand side of the nuclear burning equation, the specific energy generation rate, and a Jacobian. \\
\texttt{DynamicEngineView} & \texttt{gridfire::engine} & Abstract type for all engine views; these are used to modify the exposed physics of some underlying engine or stack of engines and views. \\
\texttt{SingleZoneSolver} & \texttt{gridfire::solver} & Abstract solver type for any single-zone network. \\
\texttt{MultiZoneSolver} & \texttt{gridfire::solver} & Abstract solver type for any multi-zone network. \\
\texttt{PartitionFunction} & \texttt{gridfire::partition} & Abstract type for any partition function, used for detailed balance calculations. \\
\texttt{ScreeningModel} & \texttt{gridfire::screening} & Abstract type used to implement plasma screening prescriptions.  \\
\texttt{NetworkPolicy} & \texttt{gridfire::policy} & Abstract type used to enforce particular minimum requirements (i.e., this network must include the proton-proton chain). \\
\texttt{ReactionChain} & \texttt{gridfire::policy} & Abstract type used to represent some reaction chain which can be used with a \texttt{NetworkPolicy} to enforce the presence of that chain. \\
\texttt{Trigger} & \texttt{gridfire::trigger} & Abstract type used to detect arbitrary state changes and trigger behavior (such as repartitioning) based on them. \\
\texttt{NetworkFileParser} & \texttt{gridfire::io} & Abstract parser type to load any user-defined network files (such as \mesa-format \texttt{.net} files). \\
\hline
\end{tabular}
\caption{Major components implemented by \gridfire. Note that this is not an exhaustive list but represent only the most important subset of abstract types.}
\label{tab:strategies}
\end{table}

\section{Reaction Rates}\label{sec:design:reaction_rates}

\gridfire is primarily focused on main sequence burning, where strong nuclear
reactions are the dominate source of both abundance evolution and energy
generation. Weak nuclear reactions are however still relevant and open both
radioactive decay pathways along with mediating neutrino cooling. The \gridfire
``reaction'' module manages both strong and weak reactions through an abstract
\texttt{Reaction} interface. Each reaction must be able to report its type
(either strong or weak), its molar reaction rate, and any relevant partial
derivatives with respect to thermodynamic state variables. \gridfire then
collects these rates to build both the right hand side of Equation
\ref{eqn:burning_2} and the system Jacobian. This
collection is managed by the engine interface (\S \ref{sec:engines}).

\subsection{Strong Rates}

\gridfire uses the JINA REACLIB database \citep{Cyburt2010} to evaluate
temperature-dependent strong nuclear reaction rates on the fly. These are
presented in the form of the coefficients, $a_{i}$, of a seven parameter
analytic equation of the form (Equation 1 in \citet{Cyburt2010}):

\begin{equation}\label{eqn:REACLIB}
  \lambda = \exp\left[a_{0} + \sum_{i=1}^{5}a_{i}T_{9}^{\frac{2i - 5}{3}} + a_{6}\ln{T_{9}}\right]
\end{equation}

Where $\lambda$ is the molar reaction rate in units of
[$\text{s}^{-1}\text{cm}^{3(N-1)}\text{mol}^{1-N}$] for a reaction with $N$
reactants. Further $T_{9}$ is the temperature in units of $1\times10^{9}\text{
K}$. We retrieve all reactions from the ``default'' library in the REACLIB
snapshot
library\footnote{https://reaclib.jinaweb.org/library.php?action=viewsnapshots,
accessed on June 17th, 2025, last modification on June 24th, 2021}. We then
filtered this database such that we only maintain reactions where the atomic
number, $Z$, of all reactant species $Z \leq$ 26 --- that is to say we remove
any reaction including a reactant species with a proton number larger than that
of iron. This choice is in line with \gridfires target domain of main sequence
burning. The design of \gridfire is such that this choice may be easily relaxed
in future versions simply by re-quering REACLIB rates and not filtering on $Z$.
Further, the python library \pynucastro \citep{SmithClark2022} provides
tools for querying up-to-data REACLIB rare data. Therefore, we have included a
utility in the \gridfire repository's utils/reaclib subdirectory which allows
for new rates to be written programmatically using \pynucastro Moreover, for
each rate \gridfire is also aware of the analytic derivative without those being
provided explicitly. This is achieved through the use of auto-differentiation
(\S \ref{sec:engines:graph:AD}).

\subsection{Weak Rates}\label{sec:design:weak_rates}
Mediated by the weak nuclear force, weak nuclear reactions comprise a key set of
reactions during stellar burning stages. Weak reactions are the primary source
of neutrinos, which in solar-like stars result in an approximate $2\%$ cooling
effect \citep{Bahcall2002}, and which are a primary means of energy escape during high energy phases
of stellar evolution such as core-collapse supernovae
\citep{Haxton2013,Maltoni2016}. \gridfire includes two separate schemes for
handling weak reactions. By default \gridfire uses the REACLIB tabulated weak
rates which are presented in the same form as strong reactions. Further, we
include a comprehensive set of weak rates from the Weak Rate Library
\citep[WRL,]{Ravlic2025}.

\begin{figure}
  \centering
  \begin{subfigure}{0.475\textwidth}
    \centering
    \includegraphics[width=0.99\linewidth]{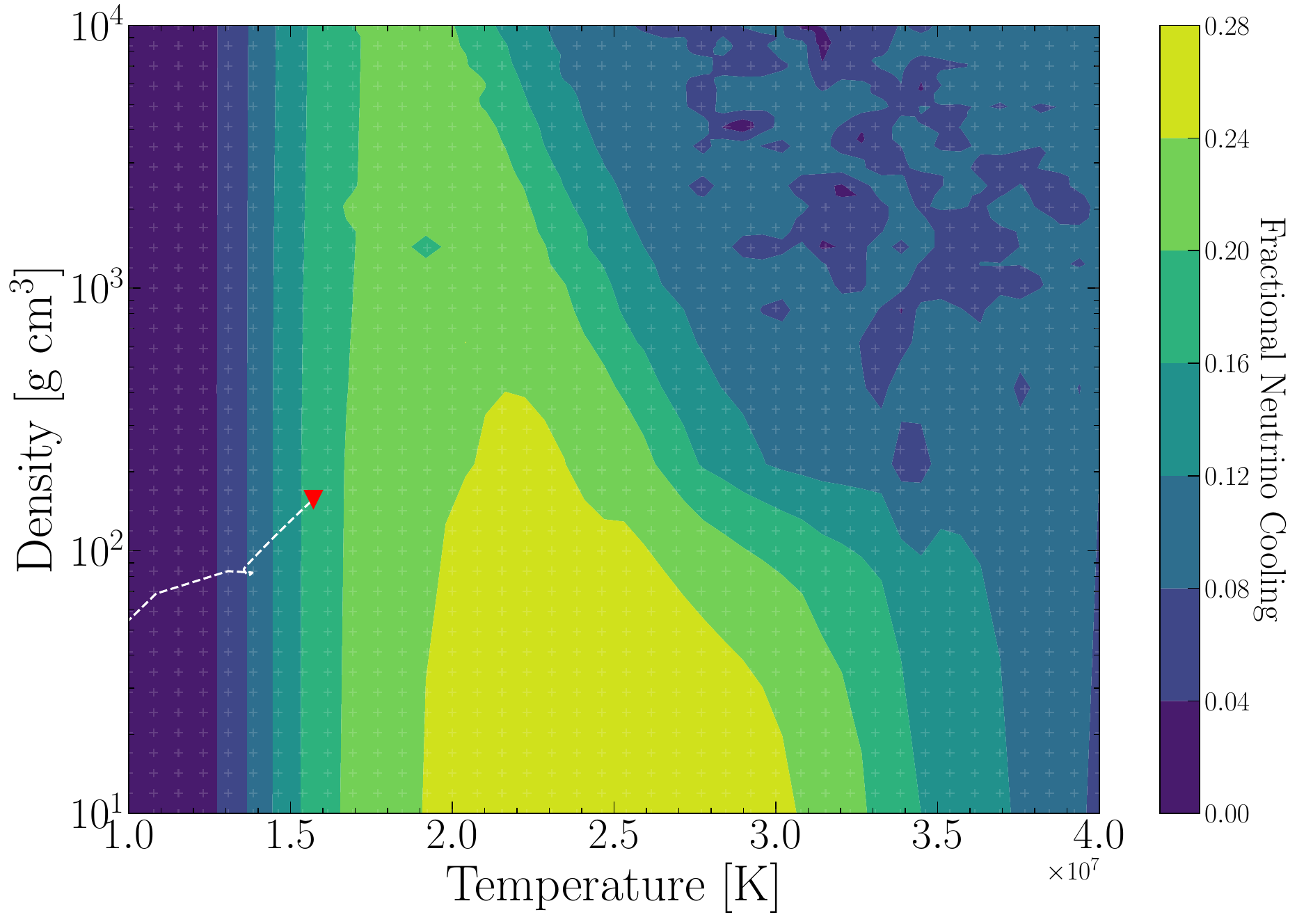}
    \caption{Ratio of total energy lost to free-streaming neutrinos on total
    nuclear energy generation after 10 years (3.1536$\times10^{8}$ s) of network
    evolution. The peak in fractional generation at $\approx T=2.2\times10^{7}K$
    corresponds to the CNO cycle not yet having reached equilibrium after only 10 years.}
    \label{fig:neutrino_cooling_A}
  \end{subfigure}
  \begin{subfigure}{0.475\textwidth}
    \centering
    \includegraphics[width=0.99\linewidth]{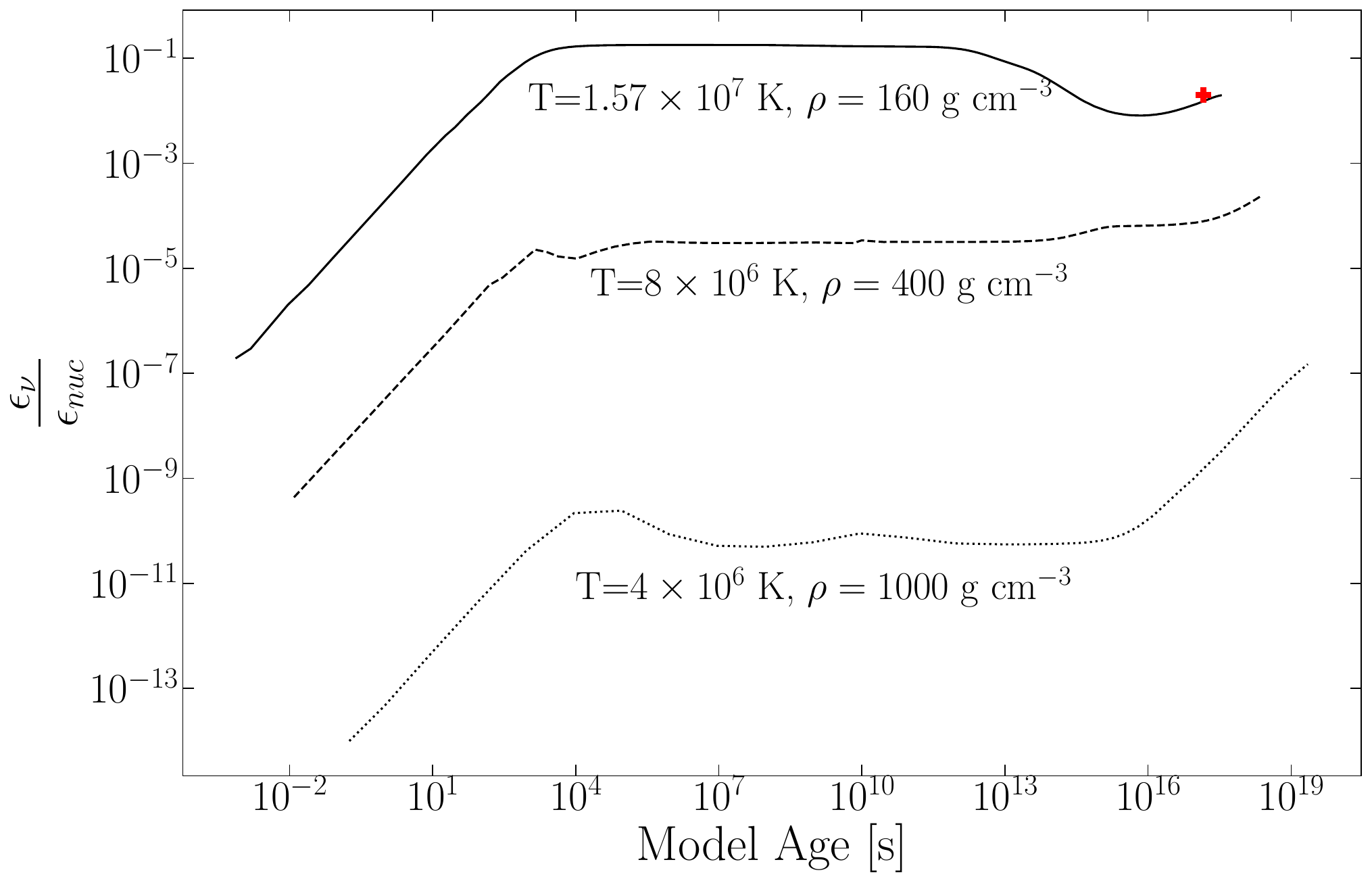}
    \caption{Ratio of total energy lost to free-streaming neutrinos on total
    nuclear energy generation for three rough approximations of core conditions
    in (from top to bottom) G-, K-, and M-dwarfs. The red + indicates the
    current day position of the Sun in this parameter space taken from
    \citet{Bahcall2002}.}
    \label{fig:neutrino_cooling_B}
  \end{subfigure}
  \caption{The neutrino cooling behavior for when \gridfire is using REACLIB
  weak reactions along with the neutrino energy loss heuristic given in Table
  \ref{tab:REACLIB_weak_summary}}
  \label{fig:neutrino_cooling}
\end{figure}

REACLIB weak rates are advantageous compared WRL rates as they may be built from
the same system we use for strong rates. Further and more importantly, as the
rate comes from the evaluation of an analytic form, derivatives are trivial to
compute using automatic differentiation. Despite these advantages, REACLIB weak
rates do not include detailed accounting of neutrino losses. Therefore, modeling
those losses when using REACLIB weak rates requires an approximate heuristic.
Table \ref{tab:REACLIB_weak_summary} summarizes this heuristic. We find that
when using this prescription \gridfire predicts an approximately $1.6\%$ cooling
effect in the solar core over 10Gyr (Figure \ref{fig:neutrino_cooling}). This is
an underestimation by $0.4\%$ compared to more accurate neutrino loss modeling
\citep{Bahcall2002}; however, in the context of main sequence modeling such a
discrepancy is acceptable.

\begin{table}
  \centering
  \begin{tabular}{c | l}
    \hline
    Type & Loss \\
    \hline
    $\beta^{+}$ & 0.5 \\
    $\beta^{-}$ & 0.5 \\
    Electron Capture & 1.0 \\
    Positron Capture & 1.0 \\
    \hline
  \end{tabular}
  \caption{Heuristic which \gridfire uses to define fractional energy lost from the mass-deficit to free-streaming neutrinos for REACLIB weak reactions.}
  \label{tab:REACLIB_weak_summary}
\end{table}

In addition to REACLIB weak rates, \gridfire also includes experimental support
for Weak Rate Library (WRL) rates. Generally these include a more robust
accounting of neutrino energy losses. However, while these are bundled with
\gridfire they are still considered experimental and are not robustly tested.
Further, as WRL rates are presented in the form of tables, including
them in the differentiation scheme requires the use of a finite difference
approximation. This imposes both a precision and runtime cost when using WRL
rates. Because of this, REACLIB rates are enabled by default, with WRL rates
disabled.  If a user would like to make use of WRL rates they may pass the flag
\texttt{WRL\_WEAK\_RATES} to the \graphengine constructor. This flag will result
in all REACLIB weak rates being excluded and all WRL weak rates being included
in the network topology. We expect that a future release of \gridfire, once
performance issues have been addressed, will change the default weak rate source
from REACLIB to WRL; further, this change may resolve the 0.4\% underestimation
of the neutrino cooling effect.

\subsection{Reverse Rates}\label{sec:design:reverse_rates} 
At temperatures above $\approx 10^{9}$ K nuclear reaction products begin to
undergo photodisintegration to a statistically significant degree
\citep[e.g.][]{Thielemann1998,Utsunomiya2006,Rauscher2018}. A proper handling of
this requires constraints on the reverse reaction rates, that is the rate of
photodisintegration. These can be either tabulated or computed using the
principal of detailed balance \citep{Klein1955, Burbidge1957,
Driller1979,Misch2017}. Reverse reactions in \gridfire use detailed balance
approach; however, these are still considered experimental and as such no work in this article includes reverse rates. The net molar
reaction flow is then the difference between the forward and reverse flow.

Detailed balance calculations require careful handling of energy level
population statistics, \gridfire uses the Rauscher-Thielmann \citep[RT,][]{Rauscher2000} 
partition function. As this is a pre-tabulated, as opposed to analytic, partition function,
auto-differentiation is not trivially possible. Rather, we implement an atomic operator 
which estimates the derivatives of the partition function using a standard finite
difference scheme. One effect of this is that, when using reverse reactions, \gridfires
performance suffers due to the need to constantly recalculate approximate derivatives. As
the target domain for \gridfire is main sequence burning we therefore choose to keep
reverse reactions disabled by default. A future paper in this series will expand the
capabilities and testing of \gridfires reverse reactions and use those to explore 
higher energy regimes where photodisintegration becomes relevant. 

\section{Engines}\label{sec:engines}
\gridfire strictly separates physics from numerical solvers. Engines (in the
\texttt{engine} module) perform all physical calculations. Primarily this looks
like computing the right hand side of Equation \ref{eqn:burning_1} along with
the Jacobian --- describing the coupling of both species abundance rates of
change to the abundance of other species as well as the rate of specific energy
generation to the abundance of all network species (Equation
\ref{eqn:JacStructure}) --- of the network.  Generally, any object which can
provide these quantities may serve as a valid \gridfire engine. Section
\ref{sec:solvers} details how these engines are stepped through time using
implicit integrators. 

\begin{equation}\label{eqn:JacStructure}
\renewcommand{\arraystretch}{1.5}
J = \begin{bmatrix}
  \frac{\partial \dot{Y}_{0}}{\partial Y_{0}} & \frac{\partial \dot{Y}_{0}}{\partial Y_{1}} & \cdots & \frac{\partial \dot{Y}_{0}}{dY_{N}}\\
  \frac{\partial \dot{Y}_{1}}{\partial Y_{0}} & \frac{\partial \dot{Y}_{1}}{\partial Y_{1}} & \cdots & \frac{\partial \dot{Y}_{1}}{dY_{N}} \\
  \vdots & \vdots & \ddots & \vdots  \\
  \frac{\partial \dot{Y}_{N}}{\partial Y_{0}} & \frac{\partial \dot{Y}_{N}}{\partial Y_{1}} & \cdots & \frac{\partial \dot{Y}_{N}}{dY_{N}} \\
  \frac{\partial \dot{\epsilon}_{nuc}}{\partial Y_{0}} &\frac{\partial \dot{\epsilon}_{nuc}}{\partial Y_{1}} & \cdots & \frac{\partial \dot{\epsilon}_{nuc}}{dY_{N}} \\
\end{bmatrix}
\end{equation}

There are two primary categories of engines in \gridfire,
\texttt{DynamicEngines} and \texttt{DynamicEngineViews}. Note that the set of all
\texttt{DynamicEngineViews} is a strict subset of the set of all \texttt{DynamicEngines}. The
separation between these categories is that \texttt{DynamicEngines} are
responsible for all of their own physics calculations, they must be fully self
contained; \texttt{DynamicEngineViews} on the other hand, may sit atop
\texttt{DynamicEngines} to form an ``engine stack''. Engines stacks provide a
clean way to adjust network structure and add new physics without the need for a
user to maintain a complete picture of \gridfires structure in their head. The
bottom of every engine stack must be a \texttt{DynamicEngine}, above that may be
stacked an arbitrary number of views.  When a caller requests some bit of
physics, say the Jacobian matrix, they will do so against the top-most view in the
stack (Figure \ref{fig:views}). This view will call the view below it, and that view calls the view below
it, and so on all the way down to the base engine. Those results then bubble
back up to the user. At each layer the view may choose to modify the results it
has received from lower layers in some way before passing them back up the stack.
A simple example might be that a user wants to model a fluid with unlimited
hydrogen, a view might be introduced which zeros out the time-derivative of the
hydrogen molar abundance thus preventing that species from time evolving. A more
complex and realistic use cases involve applying quasi-static equilibrium
prescriptions may be found in Section \ref{sec:engines:views}.
The engine system has been heavily documented and we expect that both the
\gridfire core developers and community will develop additional engine views as
\gridfire matures.

\subsection{GraphEngine}\label{sec:engines:graph}
The core dynamic engine of \gridfire is the \graphengine. This class is where
\gridfire constructs and stores its network topology. \graphengine implements
routines to evaluate Equation \ref{eqn:burning_1} and build the Jacobian matrix
(Equation \ref{eqn:JacStructure}) from REACLIB, weak rate library, and reverse
reactions. Regardless of the reaction type (details on selection and
determination of type may be found in the \gridfire documentation) \graphengine
will always calculate the net molar reaction flow for a reaction,
$\mathcal{R}_{r}$. For some reaction $r$ with reaction rate
$\lambda_{r}(T,\rho,\mathbf{Y})$, plasma screening factor $e^{H_{r}}$, symmetry
factor $\xi_{r}$ (Equation \ref{eqn:graphengine:symmetry}), abundance product
$\mathcal{Y}_{r}$ (Equation \ref{eqn:graphengine:abundance}), mass density
$\rho$, and number of reactants $N_{r}$ the forward molar reaction flow is given
by Equation \ref{eqn:graphengine:molarflow}. The symmetry factor corrects for
double counting of reactions that contain more than one reactant of the same
species, while the abundance factor adjusts for the relative molar abundance of
reactants.  Once the net-molar reaction flow is known, the stoichiometric
coefficients for reaction $r$ are used to compute $\dot{\mathbf{Y}}$ per
Equation \ref{eqn:burning_2}.

\begin{equation}\label{eqn:graphengine:symmetry}
  \xi_{r} \equiv \frac{1}{\prod_{i}^{N_{r}}n_{i}!}
\end{equation}
\begin{equation}\label{eqn:graphengine:abundance}
  \mathcal{Y}_{r} \equiv \prod_{i}^{N_{r}}Y_{i}^{n_{i}}
\end{equation}
\begin{equation}\label{eqn:graphengine:molarflow}
    \mathcal{R}_{r}(T, \rho, \mathbf{Y}) = e^{H_{r}(\mathbf{Y})}\lambda_{r}(T,\rho,\mathbf{Y})\xi_{r}\mathcal{Y}_{r}\rho^{N_{r}-1} \\
\end{equation}

Note that $\xi_{r}$ is only a function of the current reaction not the current
state of the network. Further, the plasma screening correction factor is only a
function of the current abundance and not of the thermodynamic state of the
network. \graphengine can exploit the structure of Equation
\ref{eqn:graphengine:molarflow} to aggressively cache intermediate values,
resulting in an approximately fifty times run-time decrease when compared to not
caching. Verification testing has shown identical results to within machine
precision for 64-bit floating point values when comparing \graphengine with and
without precomputation. Due to the massive speedup and demonstrated accuracy of
the precomputation route, precomputation is enabled by default; however, it 
may be disabled by users. For details on how to disable precomputation please reference
the \gridfire documentation.

\subsubsection{Automatic Differentiation}\label{sec:engines:graph:AD}
Both the stability and convergence rate of implicit solvers depend strongly on
the accuracy of the Jacobian matrix used to infer the next step. Given that
REACLIB rates are parameterized as a differentiable function (Equation \ref{eqn:REACLIB}) we
might compute and program these into \gridfire by hand. However, doing so would introduce a significant
chance of author error; further, we would be limited to whatever subset of
network topologies had been hand-derived. More adaptable approaches include
symbolic-differentiation, numerical approximations, and
automatic-differentiation (AD). \gridfires adopts AD as --- unlike numerical
methods such as finite difference --- it is theoretically exact \citep{Laue2019,
Dawood2023} while being much more performant when compared to
symbolic-differentiation \citep{Laue2019}. AD achieves its theoretically exact
results by recording the set of simple arithmetic steps which a computer
performs and then repeatedly applying the chain rule to derive an analytic derivative
for that set of steps.

Specifically \gridfire uses the C++ library \texttt{CppAD} to perform AD
\citep{Bell2025}. \texttt{CppAD} is a well-tested, tape-based, and
compiler-agnostic AD library providing both forward and
reverse mode differentiation. Derivative calculations are performed for all
functions which affect the output composition and specific energy generation;
primarily, these are the calculation of molar reaction flow per reaction and
mass deficit per reaction. The process of recording the set of arithmetic steps
is known as ``taping'' and is quite computationally expensive (taping must be
completed for all terms in Equation \ref{eqn:graphengine:molarflow} which depend
on $\mathbf{Y}$). Fortunately, taping produces a general structure rather than
any specific value, this means that we tape the network topology rather than the
current thermodynamic or composition state. \graphengine handles all of
\gridfires taping and records one tape at startup and potentially a new tape
every time the \texttt{project} function is called, if and only if the network
topology changes.  Note however that due to \gridfires view-based architecture
(\S \ref{sec:engines:views}) the underlying \graphengine topology will very
rarely change; rather, higher views can adjust how the topology is reported in a
way that does not interfere with the true topology. Thus in most instances a
tape is only recorded once at engine startup.

\subsection{EngineViews}\label{sec:engines:views}
\begin{figure}
  \centering
  \resizebox{0.95\linewidth}{!}{\begin{tikzpicture}[
    node distance=0.8cm and 1.0cm,
    font=\sffamily\footnotesize,
    >=Latex,
    startstop/.style={
        rectangle, rounded corners=3mm, minimum width=3cm, minimum height=0.8cm,
        text centered, draw=black!80, fill=red!10, thick
    },
    process/.style={
        rectangle, minimum width=3.5cm, minimum height=0.8cm,
        text centered, draw=black!80, fill=blue!5, thick
    },
    subroutine/.style={
        rectangle, minimum width=3.5cm, minimum height=0.8cm,
        inner xsep=4mm, 
        text centered, draw=black!80, fill=blue!15, thick,
        path picture={
            \draw[black] ([xshift=2mm]path picture bounding box.north west) -- ([xshift=2mm]path picture bounding box.south west);
            \draw[black] ([xshift=-2mm]path picture bounding box.north east) -- ([xshift=-2mm]path picture bounding box.south east);
        }
    },
    decision/.style={
        diamond, 
        aspect=1.8, 
        minimum width=1.5cm, 
        minimum height=0.8cm,
        text centered, draw=black!80, fill=green!10, inner sep=1pt
    },
    io/.style={
        trapezium, 
        trapezium left angle=80, trapezium right angle=100, 
        minimum width=0cm, 
        minimum height=0.8cm,
        text centered, draw=black!80, fill=orange!10,
        inner xsep=4mm 
    },
    container/.style={
        draw=gray!50, dashed, rounded corners, inner sep=0.5cm, fill=white, fill opacity=0.3
    },
    line/.style={draw, ->, thick, black!80},
    connector/.style={draw, ->, thick, black!80, rounded corners},
    label_text/.style={font=\scriptsize\itshape, text=gray!80, anchor=west}
]

\node (user) [startstop] {User};

\node (view_A) [process, below=1cm of user] {Engine View A};
\node (view_B) [process, below=1cm of view_A] {Engine View B};
\node (view_C) [process, below=1cm of view_B] {Engine View C};
\node (base_engine) [process, below=1cm of view_C] {Base Engine};

\draw [connector] ($(user.south) - (0.25, 0)$) -- +(0, -0.5) -- +(-1.5, -0.5) -- (view_A.north west) node [midway, left] {Asks for $\frac{dY_{i}}{dt}$};
\draw [connector] (view_A.south west) -- (view_B.north west) node [midway, left] {Asks for $\frac{dY_{i}}{dt}$};
\draw [connector] (view_B.south west) -- (view_C.north west) node [midway, left] {Asks for $\frac{dY_{i}}{dt}$};
\draw [connector] (view_C.south west) -- (base_engine.north west) node [midway, left] {Asks for $\frac{dY_{i}}{dt}$};

\draw [connector] (view_A.north east) -- +(0, 0.5) -- +(-1.5, 0.5) -- ($(user.south) + (0.25, 0)$) node [midway, right] {Report $\frac{dY_{i}}{dt}$};
\draw [connector] (view_B.north east) -- (view_A.south east) node [midway, right, text width=4cm] {Species $i$ not relevant to this node, act transparently};
\draw [connector] (view_C.north east) -- (view_B.south east) node [midway, right, text width=4cm] {Species $i$ in equilibrium due to this node, zero out $\frac{dY_{i}}{dt}$ before replying};
\draw [connector] (base_engine.north east) -- (view_C.south east) node [midway, right] {Calculate and return $\frac{dY_{i}}{dt}$};

\node (user2) [startstop, right=8cm of user] {User};

\node (multi_view) [process, below=1cm of user2, minimum width=5cm] {MultiscalePartitioningEngineView};
\node (file_view) [process, below=1cm of multi_view, minimum width=5cm] {FileDefinedEngineView};
\node (graph_engine) [process, below=1cm of file_view, minimum width=5cm] {GraphEngine};

\draw [connector] ($(user2.south) - (0.25, 0)$) -- +(0, -0.5) -- +(-2.2, -0.5) -- (multi_view.north west) node [midway, left] {Asks for $\frac{dY_{^{2}H}}{dt}$};
\draw [connector] (multi_view.south west) -- (file_view.north west) ;
\draw [connector] (file_view.south west) -- (graph_engine.north west) ;

\draw [connector] (multi_view.north east) -- +(0, 0.5) -- +(-2.35, 0.5) -- ($(user2.south) + (0.25, 0)$) node [midway, right] {Set $\frac{dY_{^{2}H}}{dt} = 0$, compute abundance algebraically};
\draw [connector] (file_view.north east) -- (multi_view.south east) node [midway, right, text width=4cm] {Validate that $^{2}$H is in the active species set explicitly asked for by a user}; 
\draw [connector] (graph_engine.north east) -- (file_view.south east) node [midway, right, text width=4cm] {Evaluate $\frac{dY_{^{2}H}}{dt}$ using REACLIB rates.};

\end{tikzpicture}}
  \caption{Example engine stacks (left) prototypical stack (right) specific example of a potential H-burning stack. Read each figure from left to right, the caller starts by asking for some quantity the engine knows about, for example the time-derivative of some
  species' molar abundance. First that request is passed all the way down to the base engine (left side arrows). Next, as that result bubbles back up to the caller, each
  view decides if it should modify the result in some way, views may choose to do nothing in certain cases and simply pass that quantity up to the next view or caller.}
  \label{fig:views}
\end{figure}
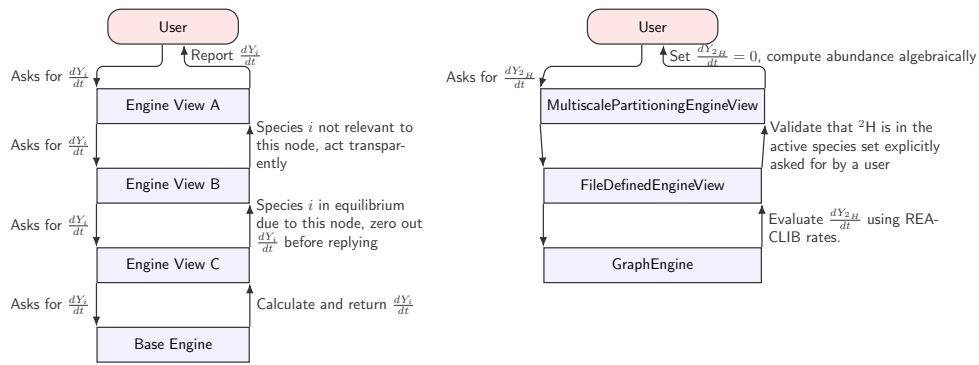

Instead of relying on wholly separate engines to adjust physics, \gridfire
adopts another standard software design pattern, the so called ``decorator''
pattern \citep[also laid out in detail in][]{Gamma1993} where ``views'' sit
atop either a base engine or one another (with a base engine always existing
at the bottom of the stack, Figure \ref{fig:views}). Each view must then implement all interface methods
from the base engine but may choose to either transparently pass arguments down
and results up or adjust those arguments and results arbitrarily. This
architecture allows for individual views to be responsible for specific physics
adjustments or simplifications and for many views to stack on top such that all
of their effects are combined. \gridfire includes a number of views and users may
implement custom or domain-specific views by following examples in the \gridfire documentation.

\subsubsection{Quasi-Static Equilibrium: \texttt{MultiscalePartitioningEngineView}}\label{sec:views:qse}
\citet{Truran1966} introduces the concept of quasi-static equilibrium (QSE),
where groups of species exchange nucleons between themselves at a much higher
rate than nucleons are leaked from the group. \citet{Hix1996, Hix1999} show that
the molar abundances of species in these groups, when solved algebraically
rather than by solving coupled differential equations, presents an accurate
approximation of the nuclear statistical equilibrium conditions found during
silicon burning and core-collapse supernova.  Further, once their equilibrium
abundances are known the time derivatives of these abundances are zero and the
species may be removed from the larger network topology. \citeauthor{Hix1996}
demonstrated that this approximation is extremely effective at loosening the
network equations during silicon burning without a significant error budget
cost. Therefore the QSE approximation may be effective at accelerating burning
for high-energy astrophysical situations. \gridfire extends this prescription
to lower-mass species and implements a robust system for automatically 
detecting equilibrium groups.

One approach to including equilibrium species / groups which has proven
successful is to generate effective reaction pathways (this approach is used by
the approx nets in \mesa). Take deuterium which in the sun, after production,
almost immediately reacts with free protons to form the much more stable
$^{3}\!\text{He}$, therefore the two reactions $\text{p} + \text{p} \rightarrow
\text{d} + e^{+}$ and $\text{p} + \text{d} \rightarrow ^{3}\!\text{He} + \gamma$
may be reduced to the effective reaction $\text{p} +\text{p} + \text{p}
\rightarrow ^{3}\!\text{He} + e^{+} + \gamma$  where the dynamic ranges of
evolutionary timescales between species is dramatically reduced. This is an
explicit statement of equilibrium. The abundance of the intermediate product ---
deuterium --- is such that the rates of both reactions are the same. 

Rather than hardcoding specific equilibrium cases, we use the
\citeauthor{Hix1999} prescriptions for QSE groups to automatically detect which
species and reactions are in equilibrium, this allows for trivial extension of
the network to new species without needed to manually calibrate equilibrium
states. These prescriptions are implemented in the
\texttt{\seqsplit{MultiscalePartitioningEngineView}}. When used the
\texttt{MultiscalePartitioningEngineView} will detect which species are in
equilibrium, set their derivatives to 0, and solve their abundances
algebraically. 

The first stage in solving for QSE abundances is to partition the network into a
slow, dynamic, set of species and multiple fast strongly connected groups. The
dynamic set comprises the species whose abundances will remain as free
parameters whereas fast species will be solved algebraically. The partitioning
algorithm\footnote{\textbf{N.B.} Partitioning takes place every time the
\texttt{project} method is called.} runs in 8 stages (A rough cartoon of the
projection process is presented in Figure \ref{fig:qseev_control_flow})

\begin{enumerate}
  \item \textbf{Network Priming:} Run a short, $1\times10^{-15}$ s at T=$10^{7}$
  K and $\rho=10^{2}$ g cm$^{-3}$ ignition to populate non-zero abundances for
  any product species. This is essential for numerically stabilizing the network
  and avoiding singularities. We have done extensive testing and such a short
  and low temperature ignition event does not affect the abundance of non-zero
  abundance species above the level of machine precision.
  \item \textbf{Timescale Based Partitioning:} Identify groups of species whose
  timescales are separated by more than two orders of magnitude. The pool which is
  slowest forms the basis of the dynamic set. Species in all other pools are
  candidate fast species.
  \item \textbf{Connectivity Analysis:} For each timescale pool identify
  connected subsets of species within the pool using a breadth-first search.
  \item \textbf{Seed Identification:} For each connected pool, identify which
  dynamic species feed and siphon mass from the equilibrium group.
  \item \textbf{Validation I:} Validate that the identified group exchanges
  nucleons internally at least 5x more strongly than it exchanges nucleons with
  non-group members. The heuristic of 5x has been chosen to
  recover equilibrium groups during main sequence burning. Researchers should
  validate this heuristic for their regimes of interest. Reference the \gridfire
  documentation for instructions on how to adjust this value.
  \item \textbf{Pruning:} Prune any reactions whose log abundance normalized
  molar reaction flow (the ratio of the molar flow to the average molar
  abundance of all species in the group, i.e. $\log_{10}(RN/\sum_{i}^{N}Y_{i})$
  where $R$ is the molar flow of some reaction and $N$ is the number of species
  in the group) is less than -30. 
  \item \textbf{Validation II:} Validate that the pruned group still exchanges
  nucleons internally at least 5x more strongly than it exchanges nucleons with
  non-group members.
  \item \textbf{Merging:} Merge any groups which are strongly coupled to each
  other
\end{enumerate}

At the end of partitioning \gridfire is left with a set of species which are not
in quasi-static equilibrium and a set of groups of species which are in
equilibrium. Once partitioning is complete the abundances for species in each
group can be solved simultaneously.  Recall that the definition of a species in
equilibrium is that the time derivative of its abundance is equal to 0,
therefore we must solve Equation \ref{eqn:QSE_1}.

\begin{equation}\label{eqn:QSE_1}
  \frac{d\mathbf{Y}_{alg}}{dt} = 0
\end{equation}

The time derivative of molar abundance is the difference in production, $P$ and
destruction, $D$, rates at a given composition, $\mathbf{Y} = [\mathbf{Y}_{alg},
\mathbf{Y}_{seed}]$, temperature, and density.  Therefore, for each species,
$i$ in the algebraic set of a QSE group we can reformulate Equation
\ref{eqn:QSE_1} as Equation \ref{eqn:QSE_1_5}.

\begin{equation}\label{eqn:QSE_1_5}
  P_{i}(\mathbf{Y}) - D_{i}(\mathbf{Y}) = 0 
\end{equation}

Because reaction rates, for a reaction with $k$ reactants, are proportional to
$\prod_{i}^{k}Y_{i}$ this is a non-linear equation. We make use of \kinsol
\citep{gardner2022,hindmarsh200}, an advanced non-linear solver, to find some
solution $\mathbf{Y}_{alg}$ satisfying Equation \ref{eqn:QSE_1_5},
$\mathbf{Y}_{seed}$ abundances are taken as the current abundance at the start
of the time step. \kinsol uses a globalized Newton method such that at each iteration, $k$,
there is some residual vector, $\mathbf{F}$, Jacobian matrix, $\mathbf{J}$, and
update vector, $\delta\mathbf{Y}$ (Equation \ref{eqn:QSE_2}).

\begin{equation}\label{eqn:QSE_2}
  \begin{aligned}
    \mathbf{F} &\equiv \left(\frac{d\mathbf{Y}}{dt}\right)^{(k)} \\
    \mathbf{J}_{ij} &\equiv \frac{\partial \left(dY_{i}/dt\right) }{\partial Y_{j}} \\
    \mathbf{Y}^{(k+1)} &\equiv \mathbf{Y}^{(k)} + \delta\mathbf{Y}^{(k)} \\
  \end{aligned}
\end{equation}

At each iteration Equation \ref{eqn:QSE_3} is solved using LU factorization,
providing the update vector for $\mathbf{Y}_{alg}$. This is repeated until the
absolute error drops below some error. By default \gridfire sets this as
one part in $10^{8}$. Users may wish to confirm that these tolerances 
are acceptable for their regime. Instructions for adjusting both absolute
and relative tolerances may be found in the \gridfire documentation

\begin{align}\label{eqn:QSE_3}
  \mathbf{J}^{(k)}\delta\mathbf{Y}^{(k)} = -\mathbf{F}(\mathbf{Y}^{(k)})
\end{align}

\kinsol is highly optimized for limited Jacobian evaluation and solver
restarting without additional Jacobian evaluations. The result of this is that
generally, after initially stabilizing into equilibrium,
\texttt{\seqsplit{MultiscalePartitioningEngineView}} converges to an abundance solution
within two or three iterations. Further, the expensive task of reconstructing
the Jacobian is limited. We have implemented a separate but numerically
equivalent solver using the C++ library \texttt{Eigen} \citep{Guennebaud2014} and find
that \kinsol is roughly 100x more performant for this use case. 

\begin{figure}
  \centering
  \includegraphics[width=0.85\textwidth]{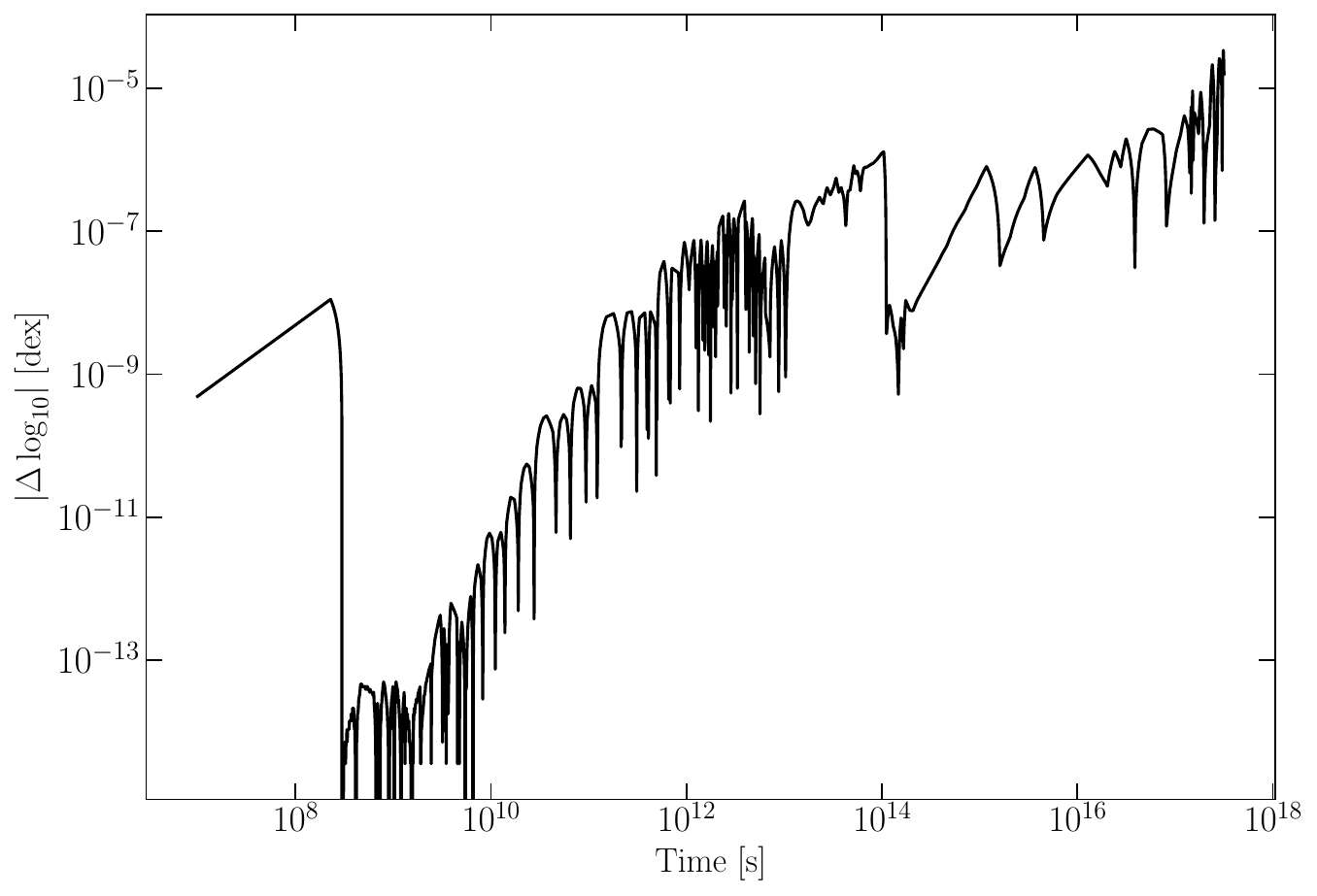}
  \caption{Logarithmic difference between D/H from an engine stack with just \texttt{GraphEngine} and a stack with \texttt{MultiscalePartitioningEngineView}. These networks
  were both evolved at solar-core like conditions of T=$1.5\times10^{7}$ K, $\rho=150$ g cm$^{-3}$ and a GS98 starting composition. The comparison was made by first interpolating 
  both runs onto the same time grid followed by a standard evaluation of the logarithmic difference.}
  \label{fig:DHErr}
\end{figure}

\begin{figure}
  \centering
  \includegraphics[width=0.85\textwidth]{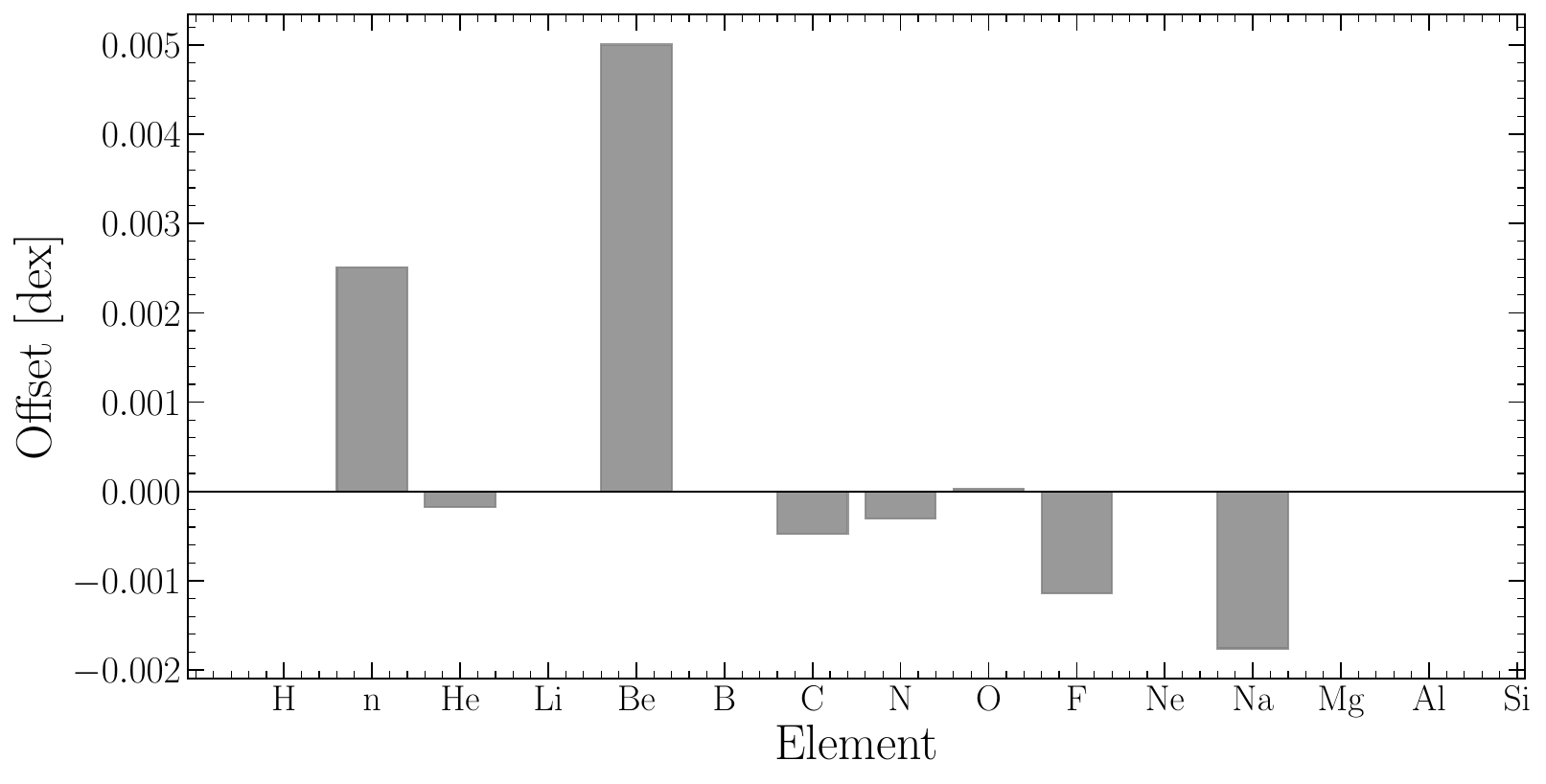}
  \caption{Logarithmic difference between elemental abundances after 10 Gyr of
  evolution from an engine stack with just \texttt{GraphEngine} and a stack with
  \texttt{MultiscalePartitioningEngineView}. Note that \gridfire does not impose
  a hard floor on species abundances; therefore, certain species have
  physically negligible trace abundances. It is trivial for these species abundances to
  vary by an order of magnitude due to floating-point round off noise which may
  spuriously be read on this figure as a $\geq$1 dex difference. We have
  therefore clipped this data to only include elements whose net abundance is
  greater than or equal to $10^{-13}$ mol g$^{-1}$ in either stack.}
  \label{fig:dexelement}
\end{figure}

Using \qsev we can make predictions about the equilibrium abundance of deuterium
in the solar-core without evolving a full network. Specifically, we find the QSE
prediction for deuterium at some time step and compare this to the proton
abundance at that time step. We find that \qsev predicts a D/H ratio
that is within $10^{-5}$ dex of the value found when integrating the full
network (Figure \ref{fig:DHErr}) and within 0.04 dex of the value predicted by
\texttt{mesa\_495.net} (see Section \ref{sec:mesa} for more details on how we
compare \gridfire to \mesa).  In general we find no species with non-trivial
abundances that differ by more than 0.005 dex between an engine stack with \qsev
and one without (Figure \ref{fig:dexelement}). Greater deviations are observed
in species with very low abundances; for example, without using \qsev \gridfire
predicts a Li molar abundance of $3\times10^{-14}$ mol g$^{-1}$ after 10 Gyr of
evolution, whereas when using \qsev \gridfire predicts a Lithium molar abundance
of $3\times 10^{-15}$ mol g$^{-1}$.  Such deviations are only seen in trace
species. Further, when using \qsev there is a maximum of a 0.004 dex offset in
specific energy and -0.002 dex in specific neutrino energy loss. While these
appear as periodic offsets, not scaling with simulation time, it is important to
remember that \gridfire is being run in a thermodynamically static environment.
Future use as part of a thermodynamically coupled system will require careful
confirmation that these errors do not become secular.

Testing thermodynamic conditions similar to the cores of M, K, F, G, A,
and B stars demonstrates that the QSE approximation introduces a maximum
relative error of one part in $10^{3}$ in species molar abundance\footnote{A
script may be found in the validation directory of the gridfire repository which
allows for more detailed exploration of these statistics
(\texttt{validation/ErrorBudget/error\_budget.py})}. Generally, unless equilibrium
approximations are particularly useful for the specific regime (e.g.  Si burning
as \citeauthor{Hix1996} demonstrate) we recommend users rely on a simpler
\texttt{GraphEngine} topology. As \gridfire is extended to cover higher energy
regimes we anticipate that \qsev will be central to maintaining effective time
stepping

\subsubsection{Other Engine Views}
In addition to the views detailed above \gridfire also includes a few utility 
views. These are \texttt{DefinedEngineView}, \texttt{FileDefinedEngineView}, and
\texttt{PrimingEngineView}. Each of these views is focused on enforcing that
\gridfire only use some limited subset of reactions. \texttt{DefinedEngineView}
allows a caller to specify the exact reaction set of interest and will only
allow network evolution along those pathways, \texttt{FileDefinedEngineView} is
the same except reading that list from a file. Finally,
\texttt{PrimingEngineView} enforces the set of reactions containing one
``priming'' species such that every reaction in that set must contain (as either
a product or reactant) that species. These views are most useful when comparing
to other nuclear networks or for numerical stabilization during early stages of
network evolution.

\section{Solvers}\label{sec:solvers}
The system of equations describing time evolution of species abundance due to
nuclear burning is extremely stiff.  Generally the stiffness of some system is
given by the ratio of the maximum to minimum real component of the absolute
value of the Eigenvalues for that system's Jacobian matrix (Equation
\ref{eqn:stiffness}). Due to the vast scale differences in the
characteristic timescales for species abundance evolution which dominate the 
spectral characteristics of the jacobian matrix, all astrophysically
interesting burning systems are pathologically stiff and therefore require implicit
solving schemes.

\begin{equation}\label{eqn:stiffness}
 S = \frac{\text{max}_{i}|\text{Re}(\lambda_{i})|}{\text{min}_{i}|\text{Re}(\lambda_{i})|}
\end{equation}

\gridfire implements two solvers, \texttt{PointSolver} and \texttt{GridSolver}.
Both solvers makes use of the \cvode library from the \texttt{SUNDIALS} suite
\citep{Cohen1996}. Specifically, we use the backwards difference formula (BDF)
family of solutions within \cvode. BDF family solvers are implicit methods which
are known to handle stiff systems extremely effectively
\citep[e.g.][]{skelboe1981, skelboe1989}. 

\subsection{\texttt{PointSolver}}
The primary solver \gridfire bundles is the \texttt{PointSolver}, a
single-zone, thermodynamically static integration scheme. \gridfire uses
\cvode's inbuilt BDF solver along with a dense linear solver
for linearized Newton steps. Each time step involves calculating the 
right hand side (RHS) of the abundance and energy evolution equations, potentially
computing the current Jacobian matrix (depending on if \cvode calls for
a recalculation of the Jacobian), a check of a ``trigger'' and, if the
trigger activates, re-projecting network state. 

\subsubsection{Right Hand Side}
The design of \gridfire calls for solvers to offload as much work as possible to
engines. Therefore, when calculating the RHS \texttt{PointSolver} is effectively
just calling the solvers required \texttt{calculateRHSandEnergy} (hereafter
$g(\textbf{x})$) member function. Note that this separation of concerns makes it
trivially easy to implement new solvers with new capabilities while reusing the
same engine. For N species $g(\textbf{x})$ has the form $g(\mathbf{x}) =
[\frac{dY_{0}}{dt},...,\frac{dY_{N}}{dt},\frac{d\epsilon_{nuc}}{dt}]$. This vector is
copied to the \cvode derivatives buffer. Details on how engines compute
$g(\textbf{x}$) may be found in Section \ref{sec:engines}.

\subsubsection{Jacobian}
Similar to $g(\mathbf{x})$ calculation, Jacobian ($J$) calculation is entirely
offloaded from the solver to the engine with the solver acting as little more
than a wrapper. The only additional step of note which the solver is responsible
for orchestrating is regularization of $J$. This is performed in two stages.
First, adjust any entries in $J$ which are less that $1\times10^{-30}$
to that floor value. Next, check if there are any $\infty$ or NaN entries in
the $J$, if not no regularization is performed, if so then stage two of
regularization begins.  Recall that $J$ has the structure from Equation
\ref{eqn:JacStructure}. In this stage the current molar abundances of both the
row and column species for each $\infty$ and NaN entry are compared to a minimum
threshold value of $10^{-100}$.  Any currently $\infty$ or NaN entries where
both abundances fall below this value are set to 0. If either species has a
current molar abundance above this cutoff then that non-finite entry remains in
the Jacobian. This will lead to an exception which is desirable behavior as a
non-finite entry in the Jacobian matrix indicates that \gridfire has strayed
into a non-physical regime and any further integration would be spurious. 

\subsubsection{Trigger \& Re-Projection}\label{sec:solvers:point:trigger}
Given \gridfires aggressive network topology optimizations (see Section
\ref{sec:engines:views}), which are fundamentally dependent on the current local
state, it is sometimes necessary to re-adjust network topology. Due to the
current topology being stored in a separate ``workspace'' chunk of memory we
call this process projection. Projection is however computationally expensive.
It may involve re-traversing the entire network graph, re-coloring the sparse
Jacobian matrix, and generally discarding most cached work and recomputing.
It is neither practical nor necessary to project every time step. Throughout much of their lives, abundances
and energy generation evolve like smooth and continuous functions and
consequently the same topology may be used over large swaths of a network's
evolutionary life. However, this is not generally true over the entire
life of a network. Major state changes, such as the exhaustion of a key fuel, 
abundance of a new fuel increasing to the point that a certain reaction begins
to contribute significantly to network evolution, or thermodynamic feedback
often necessitate reevaluation of network topology. One key question which a
solver must therefore address is ``how does one decide when to trigger a re-projection event''.

\texttt{PointSolver} addresses this through the use of \gridfires trigger system. 
Triggers are arbitrarily composable blocks of code which are provided the entire
time step context for a particular time step. \gridfires \texttt{trigger} modules 
provide standard logical combinators (e.g. AND, OR, NOT) useful when multiple
trigger conditions are required. \texttt{PointSolver} then implements four additional
conditional triggers.
\begin{enumerate}
  \item \texttt{ConvergenceFailureTrigger}: Throws when the number of convergence failures exceeds either a relative or absolute threshold.
  \item \texttt{TimestepCollapseTrigger}: Throws when the time step size falls below some relative threshold of the mean time-step over a prior window.
  \item \texttt{OffDiagonalTrigger}: Throws when any off diagonal element of the Jacobian matrix exceeds some threshold.
  \item \texttt{BoundaryFluxTrigger}: Throws when net molar reaction flow to the inactive reaction set exceeds either a relative or absolute threshold.
\end{enumerate}
These triggers are combined using logical or combinators such that if either
the number of convergence failures begins to grow, the time step begins to shrink,
the conditioning of the Jacobian becomes notably worse, or mass begins to leak into the
inactive set the solver will trigger a re-projection. 

Testing has indicated that this system is remarkably effective at recovering
during instances where the conditions provided to the \cvode are such that the
solver is unable to take steps of a practically useful size. In
solar-core like conditions when using an engine stack comprising a
\texttt{GraphEngine} and \qsev we observe that a single repartitioning event is
needed. Other regimes and engine stacks may require a different number of
re-partitioning events.

\begin{figure}
  \centering
  \includegraphics[width=\textwidth]{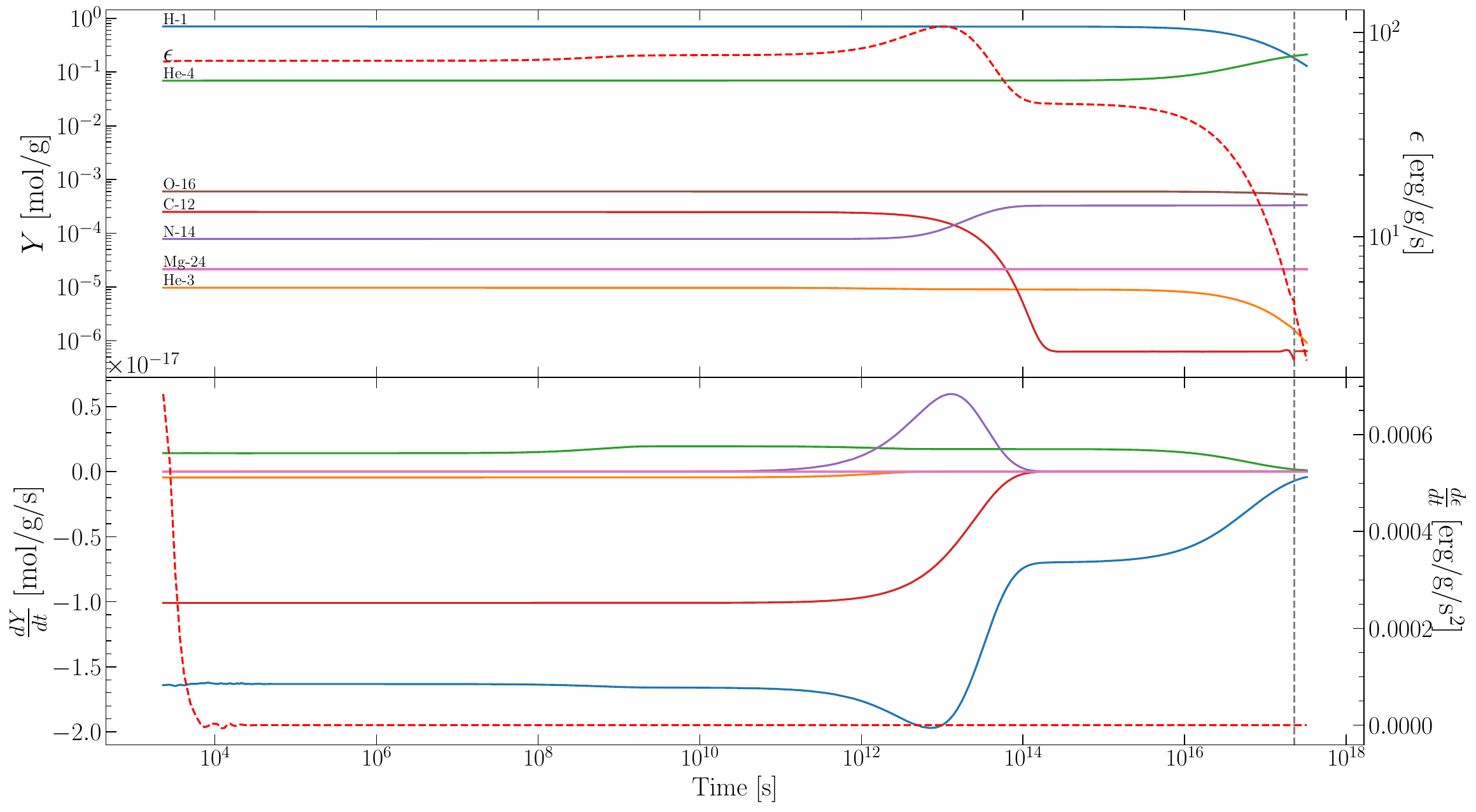}
  \caption{(upper) Plot showing species abundance evolution (solid lines) and
  specific energy (dashed line) vs time for a solar-core-like environment
  (T=$1.5\times 10^{7}$K and $\rho=160$ g cm$^{-8}$, GS98 solar composition).
  (bottom) Time derivatives of all quantities. The vertical dashed line
  represented a repartitioning event. Note that after the initial few time steps
  of stabilization the species abundances have smooth and continuous first
  derivatives. The energy generation does experience a discontinuity in
  its time derivative at repartitioning before settling back to the
  pre-repartitioning level. The impact of this discontinuity is that the
  cumulative specific energy generated after 10 Gyr deviates from the expected
  value by one part in $10^{4}$. See Section \ref{sec:solvers:point:trigger}
  for more details on how we quantify this impact.}
  \label{fig:SmoothD}
\end{figure}

One concern with any topological change is that discontinuities may be
introduced into tracked state variables. These discontinuities in fact are
problematic if we simply use reactive triggers (such as
\texttt{TimestepCollapseTrigger}) as they will sometimes throw after the network
has entered some non physical regime. For example, with just reactive triggers
and an engine stack comprising a \texttt{GraphEngine} and \qsev \gridfires
cumulative energy generation tends to not be smooth through re-projections.
These first-derivative discontinuities are explained by a slow build of mass in
the engine stack's inactive set which is then --- upon re-projection when
reactions that consume these species are enabled --- rapidly burned. The
\texttt{BoundaryFluxTrigger} resolves this by preemptively re-projecting as soon
as mass starts leaking into the engine stacks inactive set. When using this
trigger \gridfires cumulative species abundance is a mostly smooth and
continuous function. There are occasionally small discontinuities in the
specific energy generation (Figure \ref{fig:SmoothD}); however, these do not
impact the overall energy generation by an amount greater than the typical error
budget for the QSE view (\S \ref{sec:views:qse}). We quantify the impact these
discontinuities have by first identifying the time index where repartitioning
happens and deleting that point from our dataset. We use a linear interpolation
to this ``cleaned'' data. Next, we make a second copy of our original ---
uncleaned --- data, and sample the spline at the time value where the
discontinuity was. This results in two parallel datasets: one, a readout of the
simulation data, and two: a readout of the simulation data but without the
discontinuity present. We then integrate both of these datasets to find the
relative difference in the cumulative error due to the repartitioning. We find
that repartitioning introduces a relative error of $1\times 10^{-4}$. Further
testing was conducted across M, K, F, and B -like core conditions where we find
results of the same magnitude (the maximum introduced relative error being
$3\times10^{-4}$).  Finally, it should be noted that this smoothness is
primarily realized through the use of \texttt{BoundaryFluxTrigger}. We have
lightly tuned the relative and absolute tolerances of this trigger to work well
for main sequence burning.  However, end-users may wish to confirm the default
tolerances are acceptable for their regime and, if not, explore other
tolerances. Details for how to make such adjustments may be found in the
\gridfire documentation.

\subsection{GridSolver}
Generally nuclear networks in the context of SSE are required to evolve not one
region but many, each with potentially separate thermodynamic and chemical
states.  For an SSE code using the operator splitting scheme --- where nuclear
burning is handled as a separate stage than structure calculations --- a first
approach to address this might be to simply iterate over all zones of sufficient
temperature to ignite fusion, calling \texttt{PointSolver} for some time step
for each. SSE code may generally take time steps of the order of $10^{14}$s
\citep{Morel1997, Kovetz2009}.  \gridfire can, on a current generation high-end
consumer laptop, evolve an automatically optimized network with 136 species and
579 reactions over such a timescale in roughly 20 ms. For an SSE code with say
1000 burning zones\footnote{\dsep uses $\approx$100 burning zones, wherase \mesa
uses $\approx 1000$. Note however, that the comparison with \mesa is complicated
by the fact that \mesa solves its nuclear networks simultaneously with its
structure code, whereases \dsep does not; instead, taking an operator splitting
approach. } there is am immediate challenge. Namely that naive, serial,
iteration would result in a 20 second cost, per time step, just for burning.
This is a significantly larger amount of time spent on burning than most SSE
codes currently dedicate and would dramatically reduce their efficiency.

There are a few approaches which could be taken to address this challenge and
speed up multi-zone execution such as optimizing the single-zone solver more,
exploiting the embarrassingly parallel nature of this problem, or adjusting the
solver into some higher-order space to solve the entire multi-zone
network simultaneously. Significant effort has been made to increase the efficiency 
of \texttt{PointSolver} and all engine modules to a practical degree while
maintaining \gridfires goals of being easy to use and easy to develop for. In fact, the 
20 ms / 100 Myr time step quoted above is the fastest \gridfire can achieve 
given its current architecture and on current generation hardware. Therefore, we adjust the solver. 

\texttt{GridSolver} exploits the parallel nature of independent
burning. Specifically, while for $N$ burning regions $N$ local solutions to the
abundance and energy generation equations are needed, those solutions do not
depend on one another. Therefore, multi-threading may be used to compute 
some number of zones simultaneously. We find that \gridfires parallel scaling is
limited by memory-bandwidth, such that we can achieve a roughly 70\% / thread
increase in runtime  efficiency. For example, running on a computer with 16
physical cores we realize a factor of 11 runtime decrease compared to 
a naive serial execution. Given the proliferation of multi-core computers
this multithreaded approach provides a reasonable way for \gridfire to
decrease its effective runtime per burning zone. \texttt{GridSolver} then acts
as a thin wrapper around \texttt{PointSolver} and the infrastructure 
required to orchestrate parallel evaluations. Future solvers 
may choose to implement other paralyzation approaches, such as offloading 
heavy matrix operations onto Graphical Processing Units.

\subsection{Extension to Other Solvers}\label{sec:solvers:spectral}
Due to the manner in which \gridfire separates local, point-wise physics from
solvers, extending \gridfire to more complex or higher-order solver
architectures is trivial. For example \gridfires engines may be reused ---
without change --- by some new solver which couples the thermodynamics of the
system to the energy generated by fusion. Alternatively, \gridfire may be used
in conjunction with a spatial-diffusion framework to model spatially dependent
chemical evolution due to fusion. While these systems do not yet exist in
\gridfire, its architecture makes it very easy to build them. As an example we
have implemented just such a prototype demonstrating how \gridfire may be used in
conjunction with a spatial diffusion framework, \texttt{FiPy} \citep{Guyer09}, to solve multiple
zones of burning with chemical diffusion. Note that this demonstration is still run in an
thermodynamically static scheme where energy generation does not feed back into the
system temperature and density. Further, this example should be taken as a
code example of how researchers may incorporate \gridfire into their own 
tools rather than a rigorously tested scientific tool. This example file may be
found in the main \gridfire repository under the examples directory.

\section{Additional Physics}
\subsection{Plasma Screening}\label{sec:design:screening}
The bath of free electrons present in any fusing media, due to its ionization,
serves to effectively reduce the Coulomb barrier therefore increasing fusional
cross-sections \citep{Salpeter1954,Ecker1963,Stewart1966}. A number of so-called
``plasma screening'' corrections exist for various regimes of Coulomb coupling
parameter, $\Gamma \equiv \frac{(Ze)^{2}}{ak_{B}T}$ where $a$ is the average
inter-particle spacing, $Z$ is the ion-charge, $T$ is the temperature, and $e$ is the elementary charge. 
Currently \gridfire implements a weak-plasma screening model, following the
\citet{Salpeter1954} prescription, which is sufficient for \gridfires current
target domain of main-sequence burning where $\Gamma \ll 1$.

\begin{equation}\label{eqn:screening:zeta}
  \zeta \equiv 0.188\sqrt{\frac{\rho}{T_{7}^{3}}\sum_{i}^{N} \left(Z_{i}^{2} + Z_{i}\right)Y_{i}}
\end{equation}

\gridfire computes separate plasma screening factors for each reaction tracked
by the network. The process entails first computing a global,
composition-dependent, term $\zeta$ (Equation \ref{eqn:screening:zeta} for $N$
reactant species), and then for each reaction computing the screening term
$e^H$. $H$ takes a separate form depending on the number of bodies involved in
the reaction. Of astrophysical relevance are single-body reactions, two-body
reactions, and the triple-alpha process.  The forms of $H$ for these three cases
are:
\begin{itemize}
  \item Single-body: $H = 0$
  \item Two-body: $H = \zeta Z_{1} Z_{2}$
  \item Triple-$\alpha$: $H = 3\zeta Z_{\alpha}^{2}$
\end{itemize}

Once computed, the screening factor, $e^{H}$, is then a multiplicative term
applied to adjust the molar reaction flow. Because this directly affects the
mapping of an independent variable to a dependent variable, we must be able to
track the effect of this multiplication on the derivative of species abundances.
Therefore, \gridfire implements plasma screening as a auto-differentiable
module, allowing for analytic derivatives of screening factors to be
automatically computed at runtime.

Currently \gridfire only implements weak screening and bare screening (no
screening). Other regimes --- intermediate and strong screening --- are not as
directly relevant for main sequence burning. However, \gridfire is architected
in such a way as to facilitate the future incorporation of additional screening
prescriptions. We expect that future releases will include both intermediate and
strong plasma screening.

\subsection{Partition Functions}\label{sec:design:partition_functions} \gridfire
includes a partition function module, currently only used for detailed balance
calculations, which allows callers to construct piecewise partition functions.
We include both the tabulated Rauscher \& Theilmann partition function
\citep[RT,][]{Rauscher2000}, supporting species with proton numbers $Z \geq 8$,
and a ground state partition function (Equation \ref{eqn:partition:ground}).
Using the composability functions of the partition module \gridfire is able to
fall back to the ground-state partition function for species where $Z<8$.

\begin{equation}\label{eqn:partition:ground}
  Z = 2S + 1
\end{equation}

\section{Validation}\label{sec:validation}
We take a two-phase approach to validation of \gridfire. First we validate
network kinetics for stellar-like environments over long timeframes. We compare
these results to established nuclear networks to ensure consistently and
physical robustness and to demonstrate \gridfires ability to be useful for 
stellar-burning stages. These comparisions are made against both
\texttt{pynucastro} and \mesas net module through BBQ's hydrostatic mode. Second, we compare
\gridfires results for a simple big-bang nucleosynthesis calculation to
literature values, demonstrating that \gridfire can operate in higher energy
regions. Note that while we show that \gridfire can evolve a simple BBN
model, it does not include photodisintegration and should therefore be taken as
a proof of concept rather than a production ready BBN model.

\subsection{Stellar Burning}\label{sec:validation:stellar}
\subsubsection{pynucastro} 
\begin{figure}
  \centering
  \includegraphics[width=0.99\linewidth]{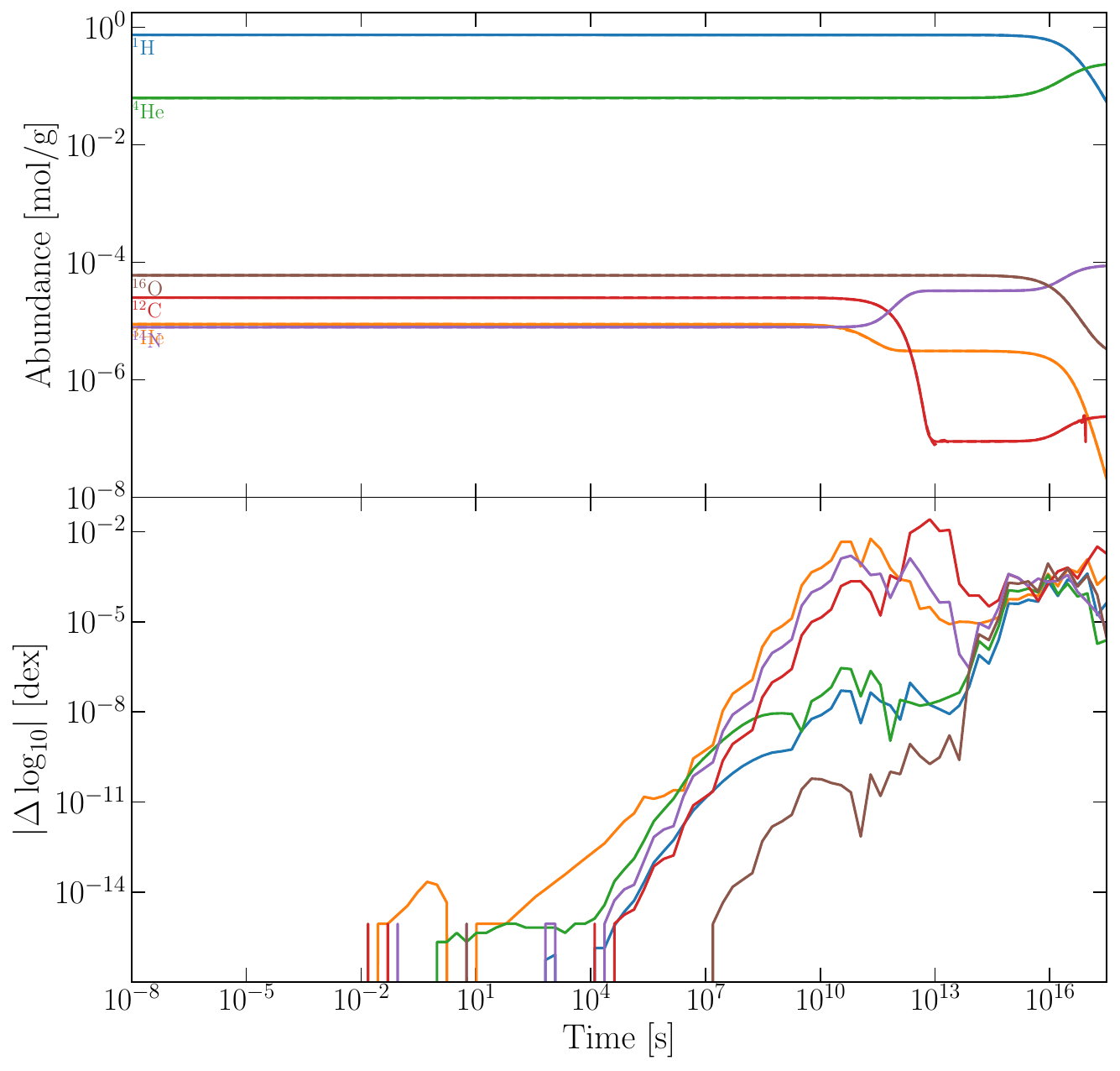}
  \caption{(upper) Molar abundances for a model evolved over 10 Gyr with a
  metal-depleted ($[Z/Z_{\odot}] = -1$) rescaling of the GS98 solar composition
  and with T=$1.5\times10^{7}$ K and $\rho=160$ g cm$^{-3}$. \gridfire results
  are plotted as solid lines, \pynucastro results are plotted as dashed lines.
  (lower) absolute value of the logarithmic difference between \gridfire molar
  abundances and \pynucastro molar abundances. This comparison was made by
  linearly interpolating both results onto the same time grid.} 
  \label{fig:vv:md}
\end{figure}

\begin{figure}
  \centering
  \includegraphics[width=0.99\linewidth]{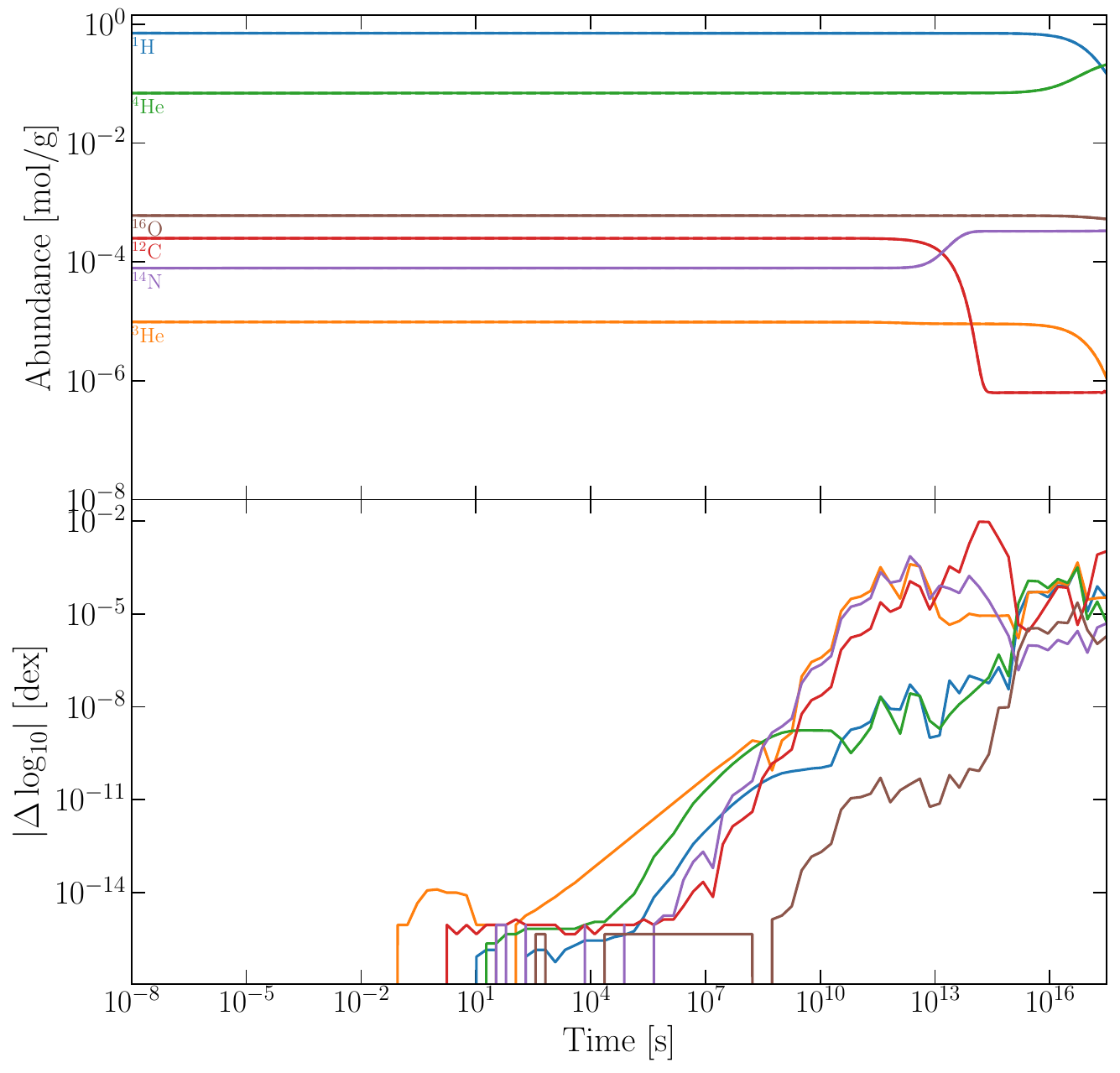}
  \caption{(upper) Molar abundances for a model evolved over 10 Gyr with a GS98
  solar composition and with T=$1.5\times10^{7}$ K and $\rho=160$ g cm$^{-3}$.
  \gridfire results are plotted as solid lines, \pynucastro results are plotted
  as dashed lines.  (lower) absolute value of the logarithmic difference between
  \gridfire molar abundances and \pynucastro molar abundances. This comparison
  was made by linearly interpolating both results onto the same time grid.} 
  \label{fig:vv:sl}
\end{figure}

\begin{figure}
  \centering
  \includegraphics[width=0.99\linewidth]{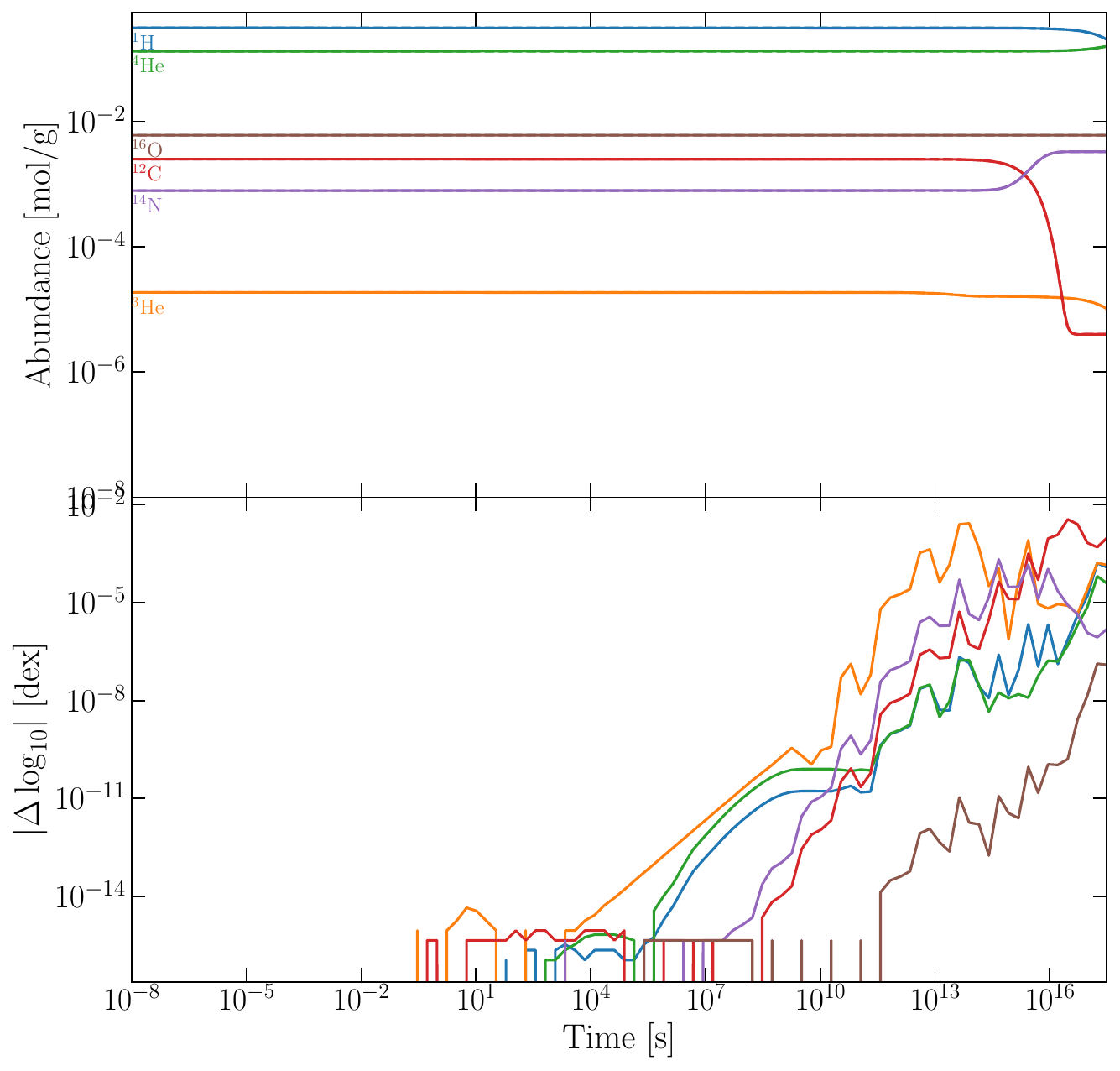}
  \caption{(upper) Molar abundances for a model evolved over 10 Gyr with a
  metal-enhanced ($[Z/Z_{\odot}] = +1$) rescaling of the GS98 solar composition
  and with T=$1.5\times10^{7}$ K and $\rho=160$ g cm$^{-3}$. \gridfire results
  are plotted as solid lines, \pynucastro results are plotted as dashed lines.
  (lower) absolute value of the logarithmic difference between \gridfire molar
  abundances and \pynucastro molar abundances. This comparison was made by
  linearly interpolating both results onto the same time grid.} 
  \label{fig:vv:me}
\end{figure}

\begin{figure}
  \centering
  \includegraphics[width=0.99\linewidth]{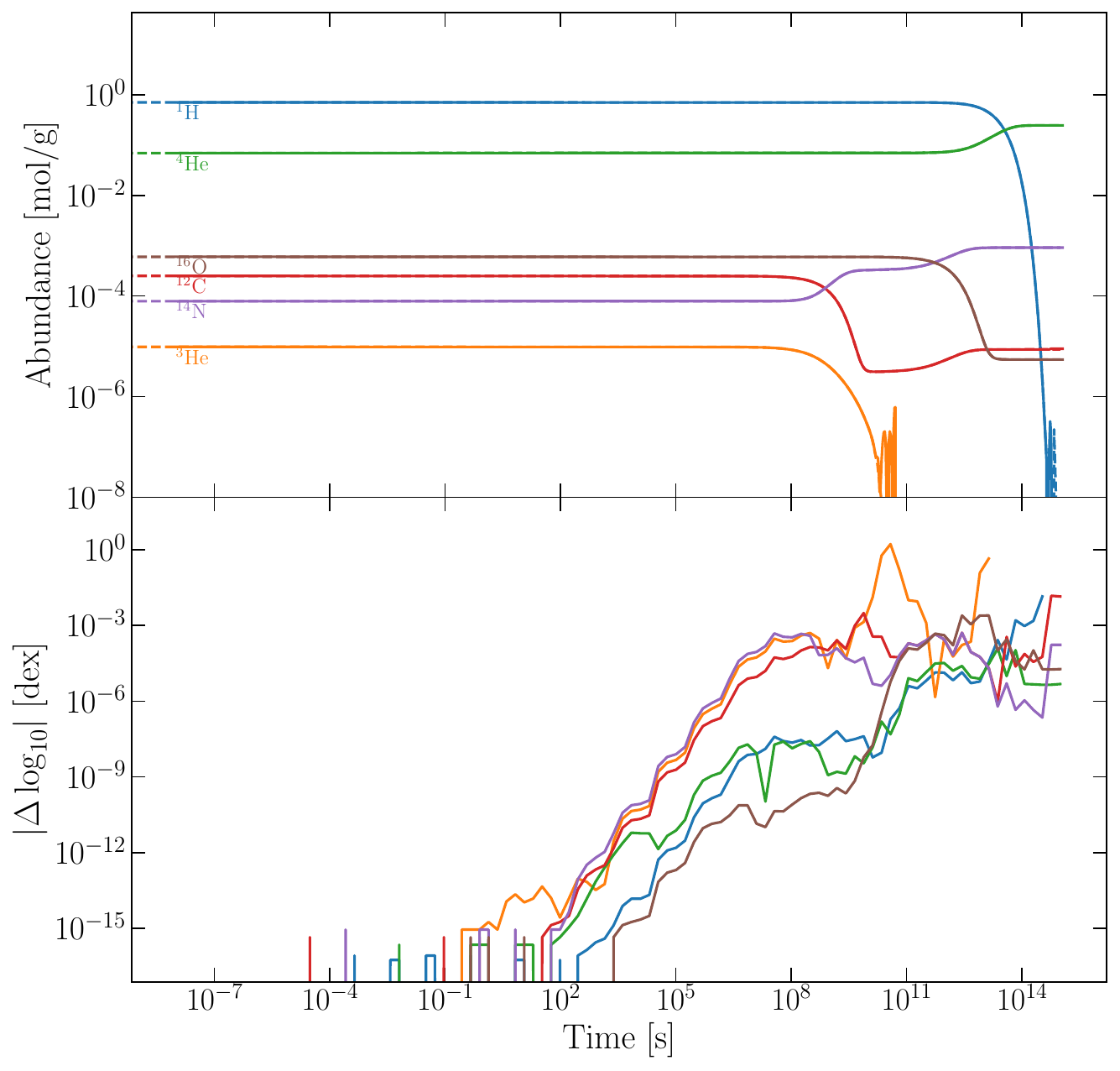}
  \caption{(upper) Molar abundances for a model evolved over 10 Gyr with a GS98 solar composition
  and with T=$4\times10^{7}$ K and $\rho=1$ g cm$^{-3}$. \gridfire results
  are plotted as solid lines, \pynucastro results are plotted as dashed lines.
  (lower) absolute value of the logarithmic difference between \gridfire molar
  abundances and \pynucastro molar abundances. This comparison was made by
  linearly interpolating both results onto the same time grid.} 
  \label{fig:vv:hs}
\end{figure}

\begin{figure}
  \centering
  \includegraphics[width=0.99\linewidth]{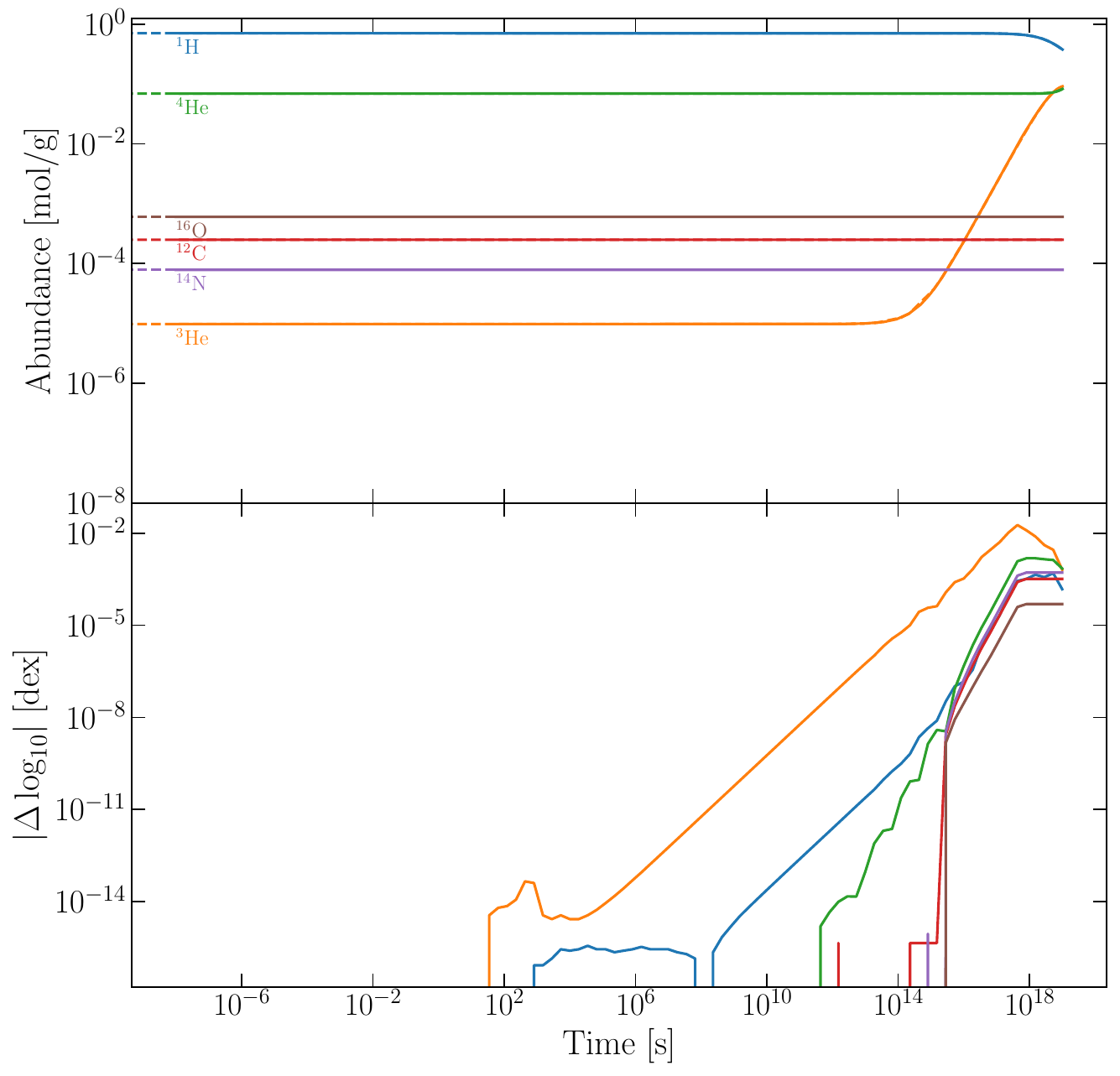}
  \caption{(upper) Molar abundances for a model evolved over 300 Gyr with a GS98 solar composition
  and with T=$4\times10^{6}$ K and $\rho=1000$ g cm$^{-3}$. \gridfire results
  are plotted as solid lines, \pynucastro results are plotted as dashed lines.
  (lower) absolute value of the logarithmic difference between \gridfire molar
  abundances and \pynucastro molar abundances. This comparison was made by
  linearly interpolating both results onto the same time grid.} 
  \label{fig:vv:cs}
\end{figure}

\begin{figure}
  \centering
  \includegraphics[width=0.95\textwidth]{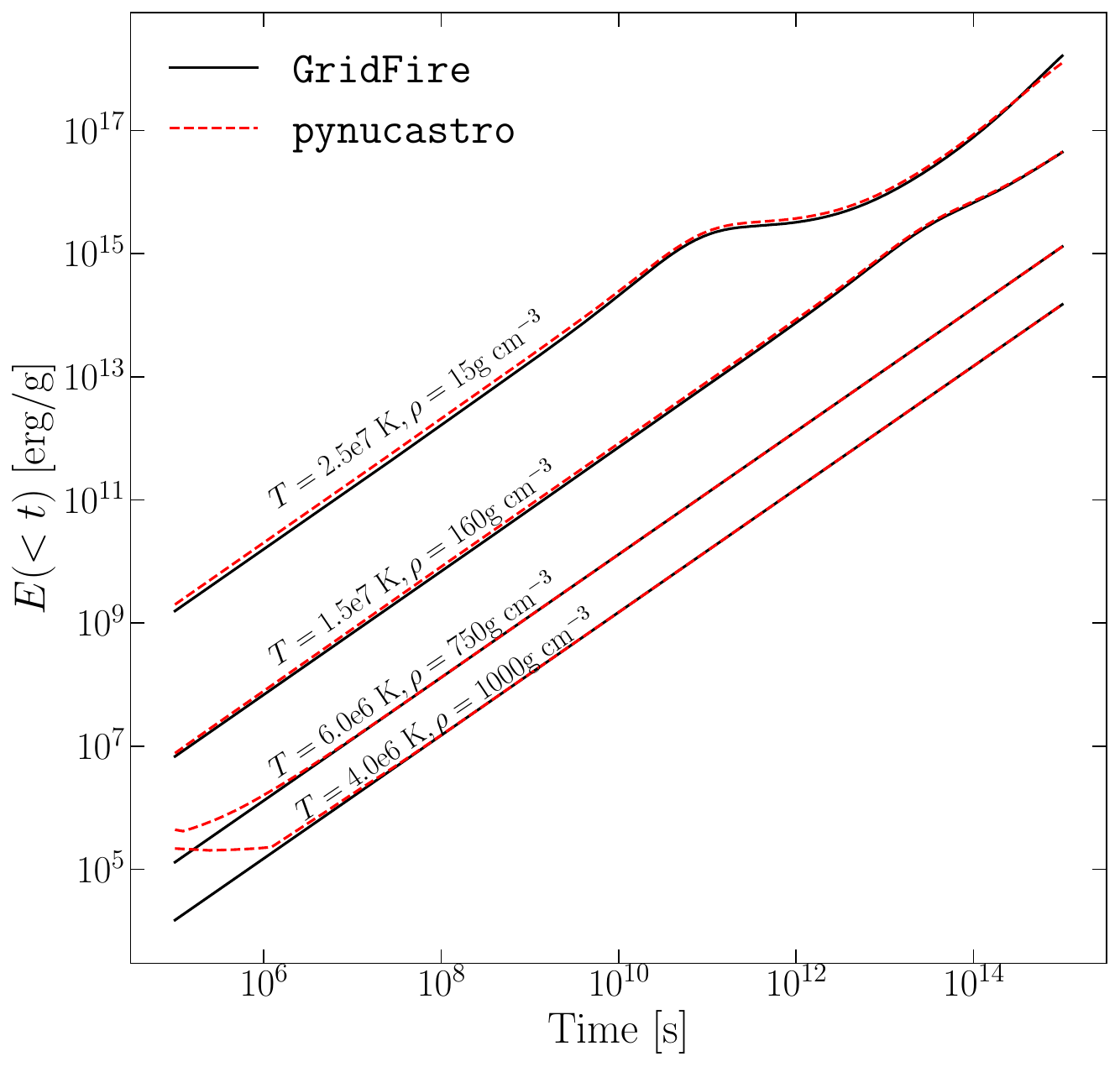}
  \caption{Cumulative specific energy generation comparison for a number
  conditions representative of stellar cores. From top-to-bottom the conditions
  roughly approximate the core of a B, G, K, M star. The initial composition is
  the same for all models, the GS98 solar-composition.}
  \label{fig:vv:eps}
\end{figure}

\gridfire results are validated against \pynucastro \citep{SmithClark2022} over
a variety of compositions and thermodynamic conditions.  \pynucastro is a well
tested, Python-based nuclear network implementation. In order to ensure
equivalent networks, we first generate the network topology using \gridfire, we
then inspect the topology and input it into \pynucastro.  Note that we read the
network topology from the base, \texttt{GraphEngine}, rather than from any
higher view. This ensures that \pynucastro is constructed from the entire
network rather than some reduced approximate form. Results from these tests are
presented in Figures \ref{fig:vv:md}, \ref{fig:vv:sl}, \ref{fig:vv:me},
\ref{fig:vv:hs}, \ref{fig:vv:cs}, and \ref{fig:vv:eps}\footnote{The
script to run these validation tests may be found in the \gridfire repository
under \texttt{validation/ManuscriptFigures/pynucastro/GridFireValidationSuite.py} Instructions for use
are included in the readme file in the validation directory.}. In summary we
find \gridfire performs well, on par with \pynucastro in both energy generation
and chemical evolution. It should be noted that while \gridfire and
\pynucastro both use REACLIB rates, the exact rate source may not be the same
between these two codes as REACLIB regularly updates and incorporates new rate
sources and measurements. Therefore, we do not expect identical results between
the two networks.


\subsubsection{\mesa \texttt{net}}\label{sec:mesa}
\begin{figure}
  \centering
  \includegraphics[width=0.95\textwidth]{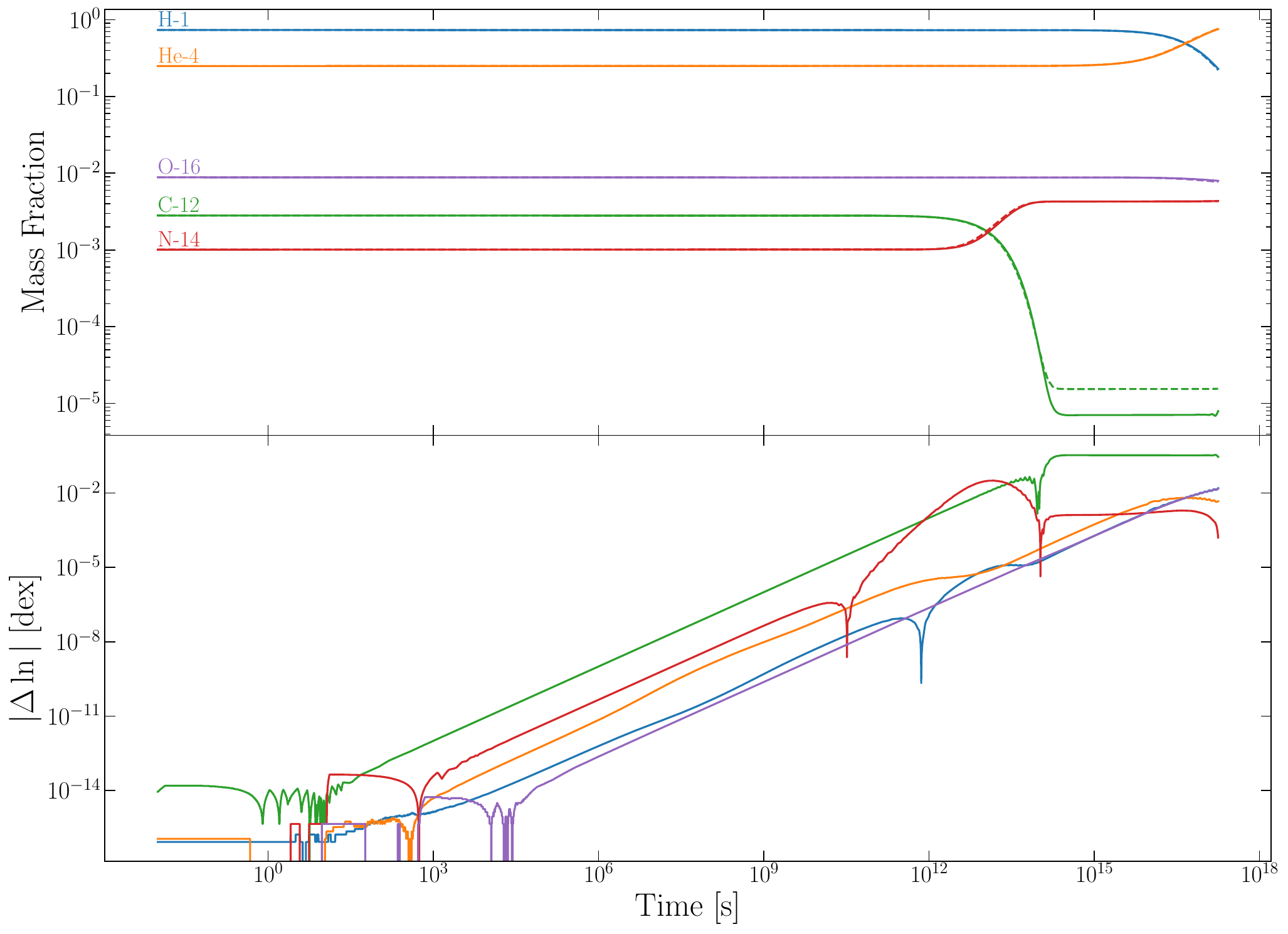}
  \caption{(upper) Comparison of Elemental Mass Fractions between \mesas \texttt{net}
  module and \gridfire. Note that all species are plotted in consistent colors.
  \mesa results are plotted as dashed lines while \gridfire results are plotted
  as solid lines. (lower) In the same color scheme as above logarithmic differences
  between each species elemental abundances. The differences between \mesa and \gridfire
  may be attributed to different underlying rate tables. These comparisons were made
  with the \texttt{mesa\_495.net} file, with an initial GS98 composition, at $T_{9}=0.015$ and $\rho=160$ g cm$^{-3}$.}
  \label{fig:GF_BBQ_Abundance}
\end{figure}

\begin{figure}
  \centering
  \includegraphics[width=0.95\textwidth]{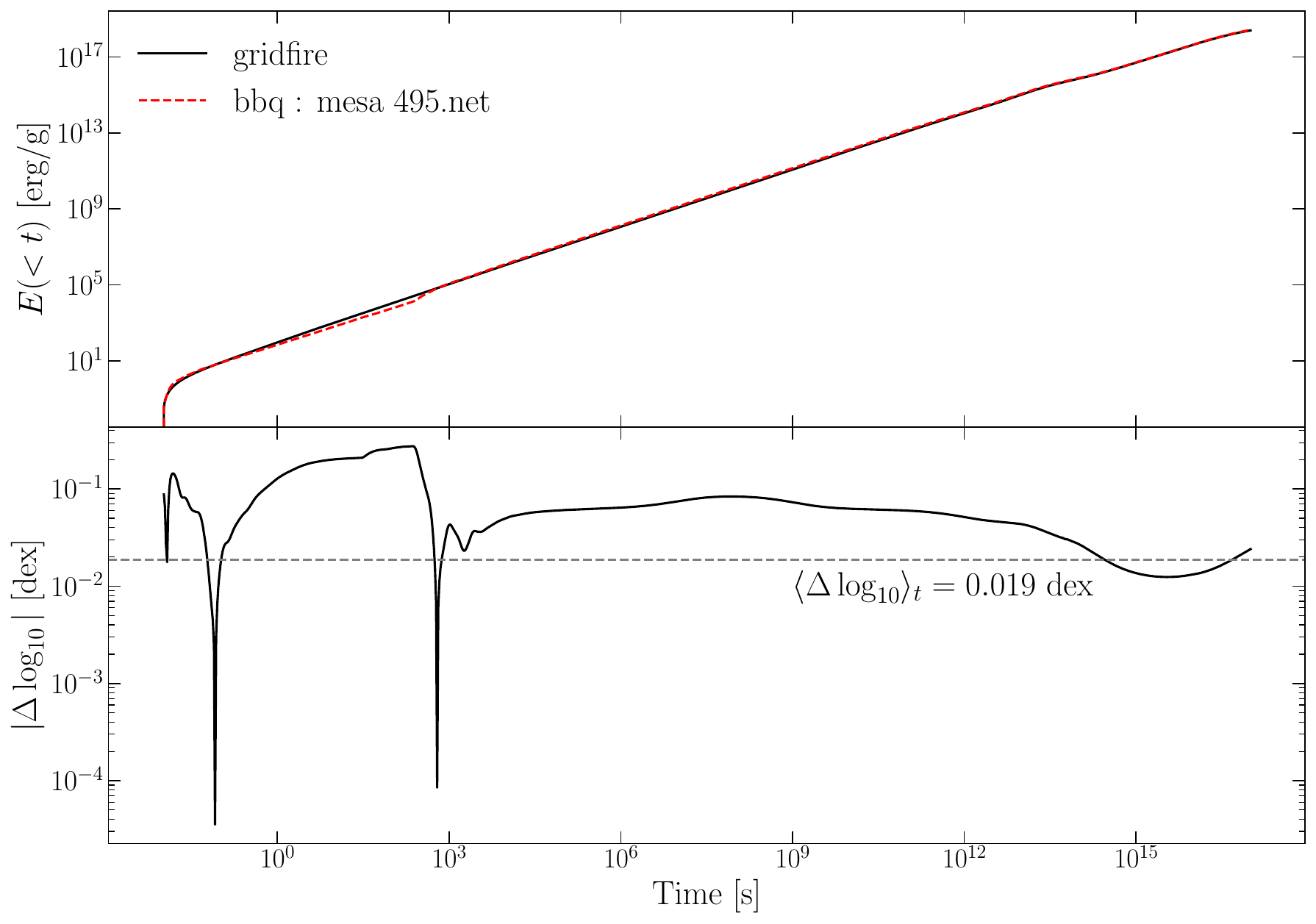}
  \caption{(upper) Cumulative Specific Energy Generation between \mesas \texttt{net}
  module and \gridfire. (lower) logarithmic difference in cumulative energy
  generation. The dashed shows the mean logarithmic difference over simulation
  time between \mesa and \gridfire. Conditions for this test are the same as described 
  in the caption of Figure \ref{fig:GF_BBQ_Abundance}.}
  \label{fig:GF_BBQ_Energy}
\end{figure}

In addition to pynucastro, we further validate \gridfire against \mesas
\texttt{net} module \citep{Paxton2011, Paxton2019}. These comparisons are made using the BBQ code \citep{bbq,Jermyn2023}
run in hydrostatic mode. This allows for abundance and energy comparisons to be
made without \mesas thermodynamic evolution adjusting plasma temperature and
density. The inlist used to run these comparisons may be found in the \gridfire
repository validation directory. All abundance and energy comparisons were run
against \texttt{mesa\_495.net}. Generally, we find good agreement between
\gridfires abundance and energy results at 10Gyr and solar-core like conditions
to those of \mesas net module (Figures \ref{fig:GF_BBQ_Abundance} \&
\ref{fig:GF_BBQ_Energy}, comparisons are made by interpolating results onto the
same time grid). The notable exception is in $^{12}C$ abundance where \gridfire
predicts a systematically lower abundance, by 0.3 dex, after the CNO cycle comes
to equilibrium compared to what \mesas \texttt{net} predicts.

This difference can be attributed to \gridfire using only rates from REACLIB
while \mesa makes uses of a more curated rate set. In the case of $^{12}C$, \citet{Formicola2004,Imbriani2005} 
show that measurements of $^{14}N(p,\gamma)^{15}O$ may be up to 2x slower at low 
energies when compared to theoretical measurements. Figure 15 in \citeauthor{Imbriani2005}
demonstrates that below $T_6 \approx 100$ the true reaction rate of $^{14}N(p,\gamma)^{15}O$
may be 0.5x that of the NACRE compilation. We can demonstrate that this discrepency 
explains the majority of the 0.3 dex offset seen between \gridfires results and \mesas results 
by considering an equilibrium condition for the CNO-I cycle. In equilibrium the rate of $^{12}C$ destruction must equal 
the rate of $^{14}N$ destruction (Equation \ref{eqn:eq:cno-1})

\begin{align}\label{eqn:eq:cno-1}
  N(^{12}C)N(H)\lambda_{12} &= N(^{14}N)N(H)\lambda_{14}
\end{align}

Where $N(^{A}X)$ is the number density of species $^{A}X$, and $\lambda$ is the reaction rate. Introducing a factor of the atomic mass 
of species allows for us to solve for the equilibrium abundances.

\begin{align}
  X(^{12}C) &= \frac{12}{14}X(^{14}N)\left(\frac{\lambda_{14}}{\lambda_{12}}\right) \\
  \log_{10}(X(^{12}C)) &\propto \log_{10}\left(X(^{14}N)\right) + \log_{10}\left(\frac{\lambda_{14}}{\lambda_{12}}\right) \\
\end{align}

The estimated logarithmic difference in carbon-12 mass fraction between
two networks, $A$ and $B$, with independent rates is then given by Equation \ref{eqn:cno:ratio}.

\begin{align}
  \Delta\log_{10}(X(^{12}C))_{AB} &= \log_{10}(X(^{12}C))_{A} - \log_{10}(X(^{12}C))_{B} \\
  \Delta\log_{10}(X(^{12}C)) &= \log_{10}\left(\frac{\lambda_{14,A}}{\lambda_{12,A}}\right) - \log_{10}\left(\frac{\lambda_{14,B}}{\lambda_{12,B}}\right) \label{eqn:cno:ratio}
\end{align}

The \texttt{rates/test/src/test\_rates.f90} tool provided by \mesa
evaluated at $T_9$=0.015 shows that \mesas \texttt{net} uses
$\lambda_{12,\text{\mesa}}=2.89\times10^{-16}$ and
$\lambda_{14,\text{\mesa}}=1.16\times10^{-18}$. Similarly, we can use the
\texttt{calculate\_rate} method on each of these reactions in \gridfire to find
$\lambda_{12,\gridfire}=3.59\times10^{-16}$ and
$\lambda_{14,\gridfire}=6.92\times10^{-19}$.  Plugging these into Equation
\ref{eqn:cno:ratio} yields a -0.32 dex difference, nearly exactly what we
observe; further, we can confirm that \gridfire uses NACRE rates by probing the
label attribute on each of these reactions. Generally then this discrepancy may
be attributed to differences in the underlying rate table rather than a structural issue with
\gridfire. In the future we may update \gridfire to include additional rate sets.

\subsection{Big Bang Nucleosynthesis}\label{sec:validation:bbn}
In the first minutes after the big bang the conditions were such that free
protons and neutrons underwent significant fusion resulting in the primordial
big bang nucleosynthethis (BBN) composition which can be observed in high
red-shift quasars \citep{Steigman2009,Pettini2012}, extragalactic HII regions
\citep{Olive1997,Cyburt2005}, and metal-poor halo stars \citep{Fields2005}. A
full handling of this requires a proper accounting for photodisintegration as
cosmological models predict that temperatures were well above the roughly 1 GK
cutoff where photodisintegration is relevant. However, the initial $\approx$
0.1 -- 180 seconds of the universe's life were dominated by neutron-proton
reactions which set some neutron-proton ratio. If we adopt a neutron-proton
ratio from after the deuterium bottleneck \citep{Turner2021} and let \gridfire evolve from there we may
validate \gridfires BBN predictions. 

Letting the baryon density $\Omega_{b}h^{2} = 0.022$
\citep{PlanckCollaboration2016} we adopt a proton-neutron ratio of $X_{p}/X_{n}
= 7.17$ \citep{Yeh2023}. We start \gridfire with a composition only composed of
neutron and protons and allow \gridfire to use its built-in topology generation
and optimization schemes to expand this as necessary. Further, since \gridfire
is currently thermodynamically static over any given integration scheme we make
use of an operator-splitting approach to time evolve the temperature of the
universe.

Following a simple Friedmann cosmology --- where the scale factor in a radiation
dominated universe goes like $a\propto t^{1/2}$ and the temperature goes like
$T\propto a^{-1}$ and with the number of flavors of light neutrinos, $N_{\nu}$ =
3 --- we let the temperature and density follow the relations given in Equation
\ref{eqn:bbn:thermo}. Further, we adopt a geometric time
stepping approach where the time step requested by our BBN program is given as
$\Delta t = t + \alpha t$ where $\alpha$ is some configurable constant
(generally we find that known transient features are resolved well with
$\alpha=0.01$). BBN is expected to have reached completion by roughly the
neutron lifetime in the early universe, $\tau_{n} \approx 886$ s
\citep{Frekers2025}, after the big bang; consequently, we evolve to 1200s to
ensure that \gridfire has enough integration time to stabilize any equilibrium
species and to ensure that we do not observe any spurious behavior after BBN is
expected to have completed.

\begin{equation}
  \begin{aligned}
    T_{9}(t) &= \frac{10}{\sqrt{t}} \\
    \rho(t) &= 4\times 10^{-5} \cdot T_{9}^{3}(t)
  \end{aligned}
  \label{eqn:bbn:thermo}
\end{equation}

\begin{table}
  \centering
  \begin{tabular}{r | c c c}
    \hline
    Species & Initial Mass Fraction & Final Mass Fraction & $X_{i}/X_{p}$ \\
    \hline
    $^{1}$H & 0.88 & 7.6147$\times 10^{-1}$ & 1 \\
    $^{2}$H & 0 & 2.8482$\times 10^{-5}$ & 3.7404$\times 10^{-5}$ \\
    $^{3}$He & 0 & 2.2286$\times 10^{-5}$ & 2.9268$\times 10^{-5}$ \\
    $^{4}$He & 0 & 2.3848$\times 10^{-1}$ & 3.1319$\times 10^{-1}$ \\
    $^{7}$Li & 0 & 1.1379$\times 10^{-10}$ & $1.4944\times 10^{-10}$ \\
    $^{7}$Be & 0 & 3.2127$\times 10^{-9}$ & $4.2191\times 10^{-9}$ \\
    \hline
  \end{tabular}
  \caption{Big Bang Nucleosynthesis (BBN) abundances as predicted by \gridfire
  after 1200s following the temperature and density profile described by
  Equation \ref{eqn:bbn:thermo}.}
  \label{tab:bbn:abundances}
\end{table}

\begin{figure}
  \centering
  \includegraphics[width=0.95\textwidth]{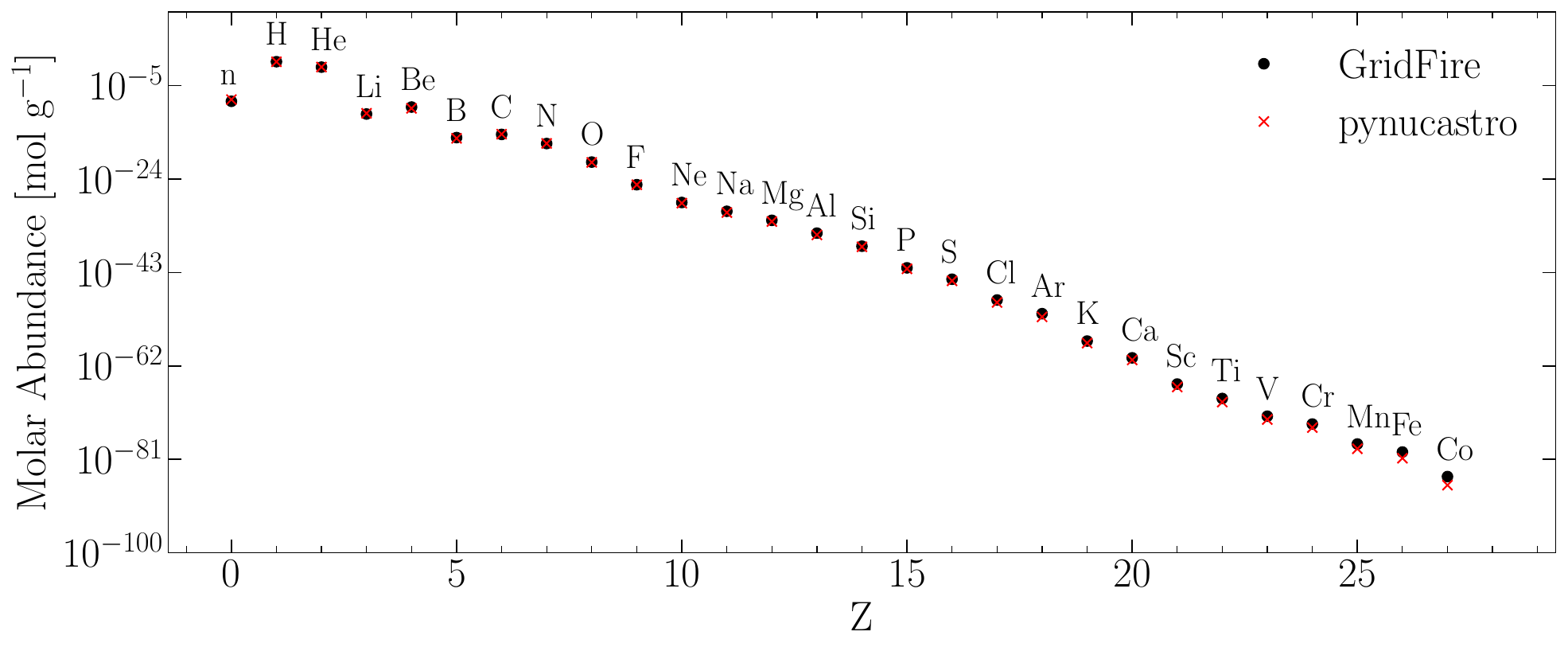}
  \caption{Elemental Abundances \gridfire predicts after 1200s of evolution
  following the temperature and density profile described by Equation
  \ref{eqn:bbn:thermo}. Similar to results presented in Section
  \ref{sec:validation:stellar} we make the pynucastro comparison by copying the
  \gridfire topology and running \pynucastro through the same operator split
  scheme. The script used to generate this data may be found in the \gridfire
  repository under validation/ManuscriptFigures/BBN.}
  \label{fig:bbn:abundance}
\end{figure}

\begin{figure}
  \centering
  \includegraphics[width=0.75\textwidth]{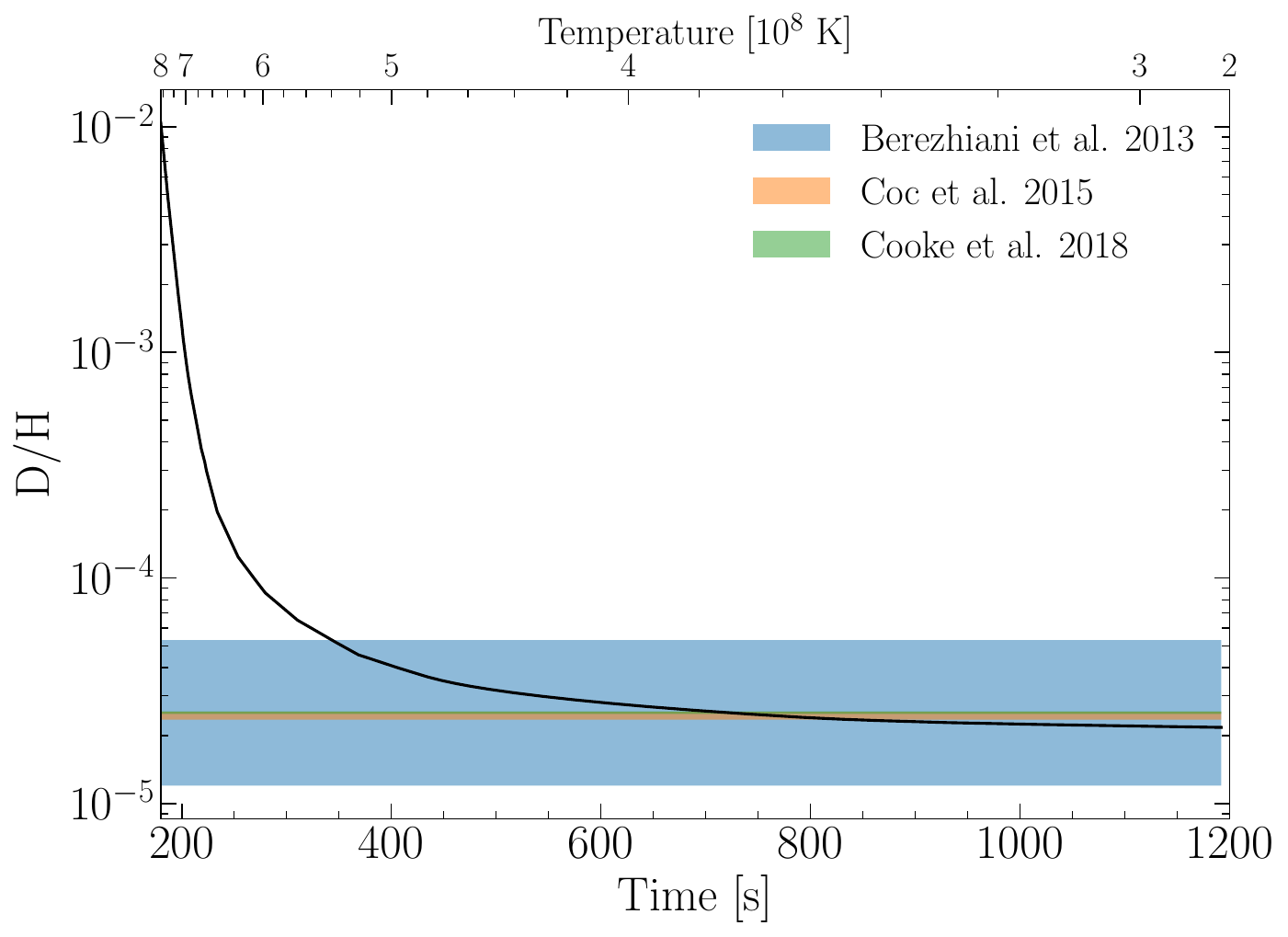}
  \caption{\gridfire deuterium on hydrogen mass fraction ratio from t=180s to
  t=1200s after the big bang. Comparisons are drawn against
  \citet{Berezhiani2013,Coc2015,cooke2018one}. \gridfire agrees to within 13\%
  of \citeauthor{cooke2018one}; differences are attributable to our
  simplistic BBN model.}
  \label{fig:bbn:dh}
\end{figure}

\begin{figure}
  \centering
  \includegraphics[width=0.75\textwidth]{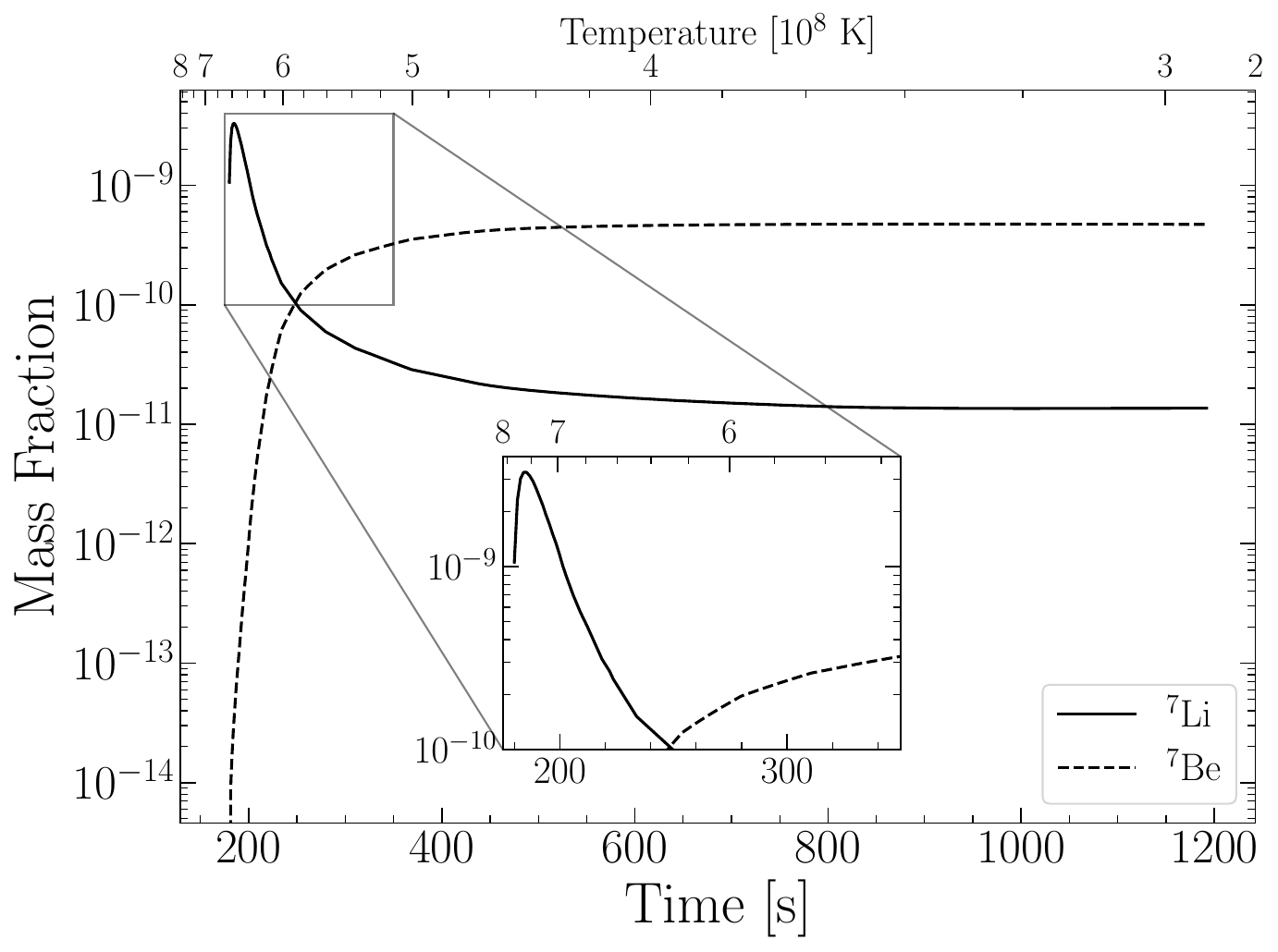}
  \caption{\gridfires predicted mass fraction evolution of $^{7}$Li and $^{7}$Be.}
  \label{fig:bbn:a7}
\end{figure}

We find the \gridfire is able to reproduce both BBN abundance values and
transient structure --- such as the temporary peak in $^{7}$Li abundance
\citep[e.g.][]{Olive2000, Mishra2011, Jang2021}--- in abundance evolution of
trace species to within approximately thirteen percent of high precision results
from \citet{cooke2018one} (Table \ref{tab:bbn:abundances}, Figures
\ref{fig:bbn:abundance} \& \ref{fig:bbn:dh}). The discrepancy here is likely due to our
simplistic BBN model. We find that $^{7}$Li and $^{7}$Be abundances match
expectations, with a peak in $^{7}$Li abundance at 185 seconds, followed by
$^{7}$Be approaching an equilibrium abundance of $\approx 10^{-9}$ (Figure
\ref{fig:bbn:a7}).

\section{Performance}\label{sec:performance}
In order to not cause a significant slow down in runtime for current SSE
programs, \gridfire must be capable of evolving a single time step's nuclear
burning calculations over a wall-time, $t_{gf}$, such that $t_{gf} \ll t_{s}$
where $t_{s}$ is the wall time to solve the structure equations. Further,
\gridfire must not increase overall memory usage such that running \gridfire and
some SSE program simultaneously would be impractical on common modern hardware.
Here we will use a one solar mass model evolved with the Dartmouth Stellar
Evolution Program \citep[\dsep,][]{Dotter2008} as a point of comparison. Our
testing indicates that for these conditions \dsep takes an average of five
seconds of wall time per 100 Myr time step\footnote{This has been measured when
evolving with \texttt{free-eos} enabled, 1000 shells, and with numerical
tolerances set to one part in $10^8$}. Therefore, we adopt a target for
\gridfire performance such that \gridfire should be able to evolve 100 burning
zones (\dsep uses 100 burning zones) in less than one second of wall time.
Readers should note that performance numbers are strongly dependent on
hardware and exact performance characteristics will vary depending on hardware.

\subsection{Compute Time}\label{sec:performance:compute_time}
\begin{figure}
  \centering
  \includegraphics[width=0.95\textwidth]{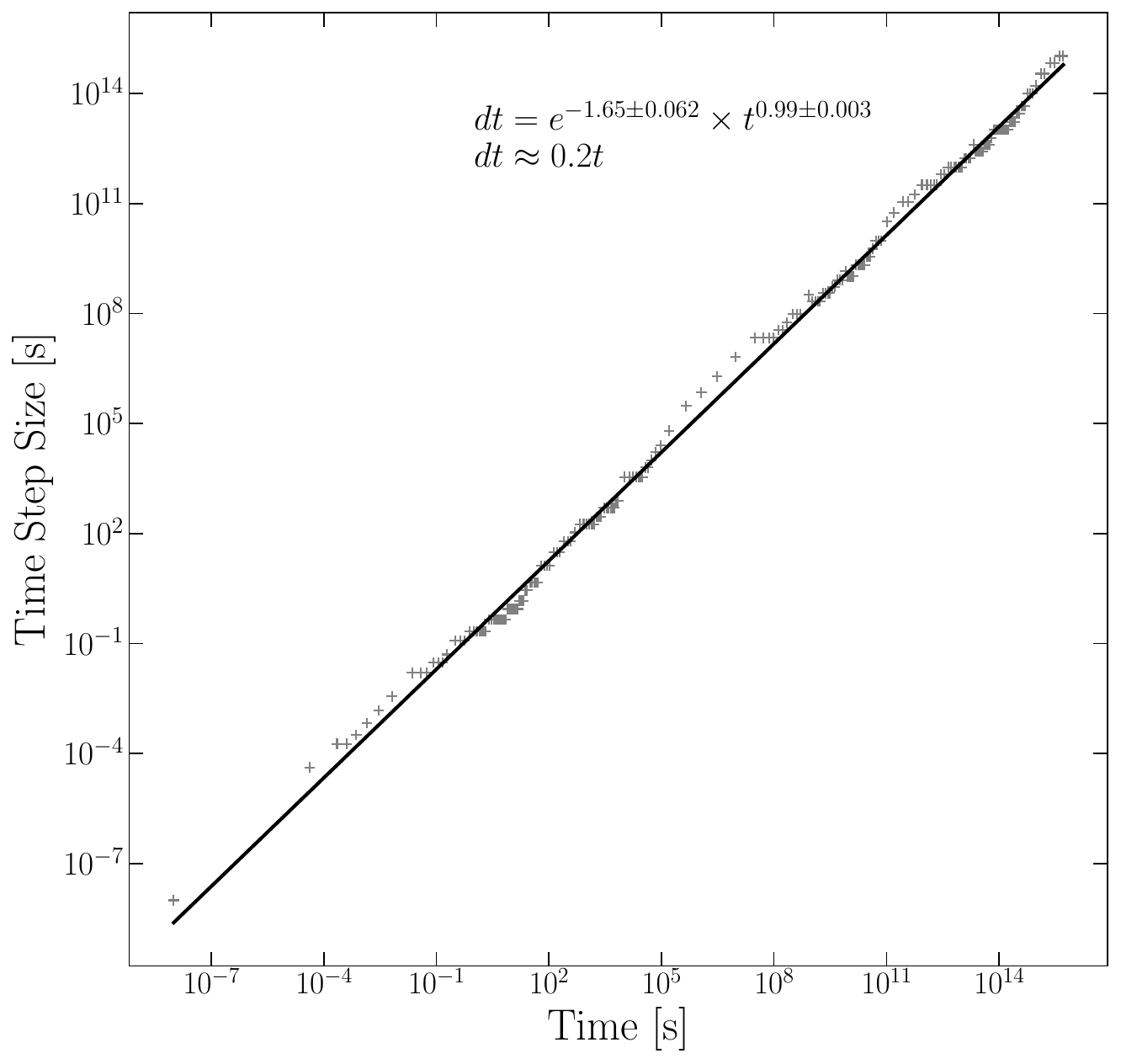}
  \caption{Time step scaling for a \gridfire run with numerical tolerances set at one part in $10^8$ for
  a network run at solar-core like conditions.}
  \label{fig:simtime_vs_walltime}
\end{figure}

\gridfires wall time per evaluation (which is the relevant quantity when
considering an operator splitting scheme) is dependent on the simulation 
length for the evaluation. That is to say that evaluating an engine stack over
10 years will take less time than evaluating that same stack over 10 Gyr. However,
the effect is not linear as \cvode is extremely effective at ramping step
sizes up as a network stabilizes such that, for solar-like stars, the time step size
scales roughly like the current simulation time (Figure \ref{fig:simtime_vs_walltime}).
Note that this generally means that the majority of wall time is spent evaluating
the early stages of network evolution. 

Further, at higher temperatures and densities an increasing fraction of
reactions may contribute to network kinetics. This results in an increased
spread in the real components of the network Jabobian's eigenvalues which in
turn forces \cvode to take smaller timesteps; consequently, the wall time
needed for \gridfire to evaluate the same simulation time generally increases
with temperature and density (Figure \ref{fig:MESA_GF_Time}). This is not a
strictly monotonic relation due to various reaction pathways entering and
exiting relevance, species being partitioning in or out of the fast set, and the
general complexity of this system of equations complicates the shape of the
parameter space.

\begin{figure}
  \centering
  \includegraphics[width=0.85\textwidth]{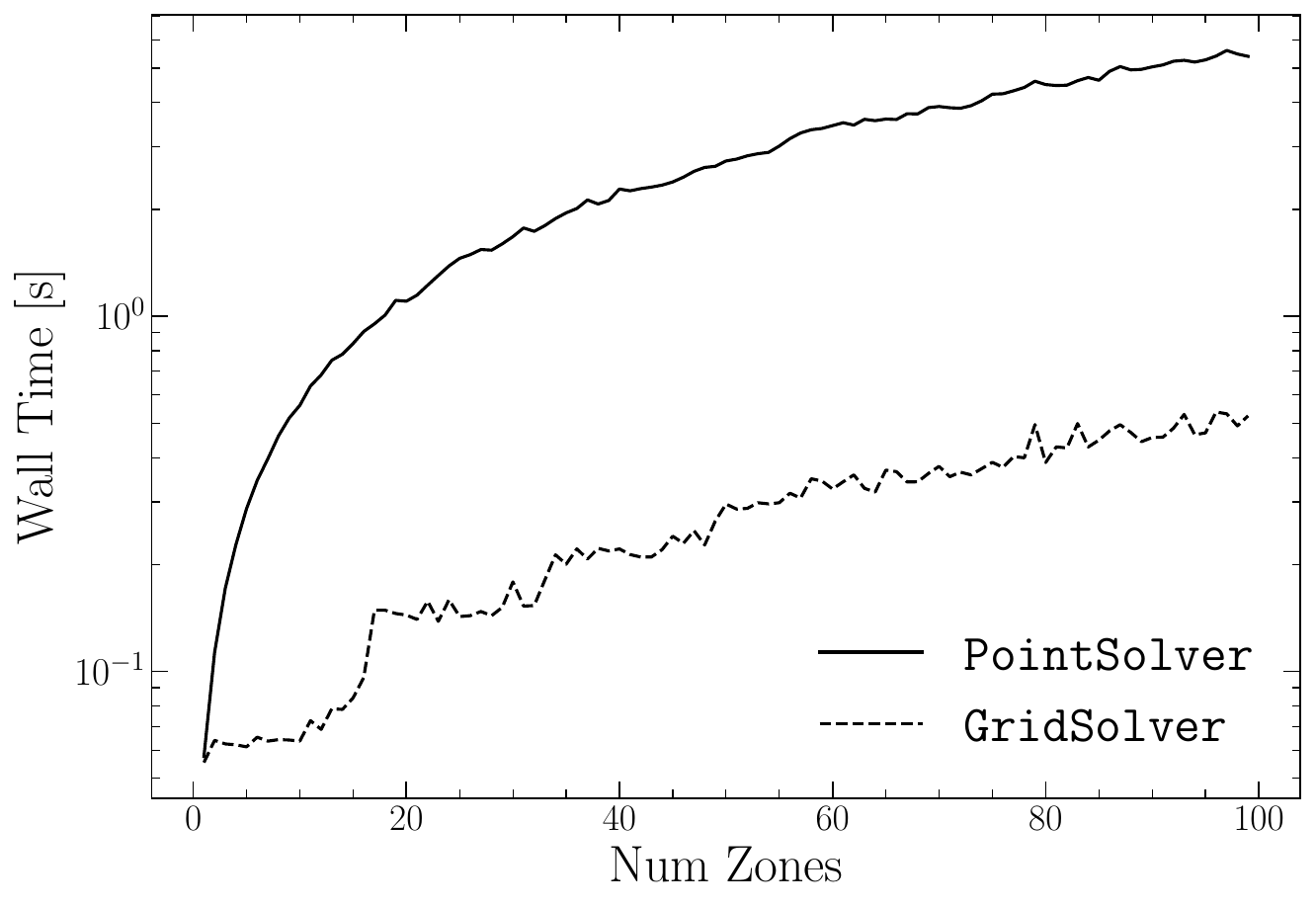}
  \caption{Mean wall time scaling for \texttt{PointSolver} and
  \texttt{GridSolver} for a solar-core-like network over 1Gyr. Each data point 
  is the mean of 1000 runs for that number of zones. That is to say that the
  wall time reported for a zone number of 20 is the mean of 1000 runs of the
  solver for that number of zones. This test was run on a 2024 M4 Max MacBook
  Pro with 16 physical cores, 12 high performance cores and 4 efficiency cores.
  Results will depend strongly on hardware. The jump in \texttt{GridSolver}'s 
  wall time at $\approx$ 16 zones corresponds to the point at which all cores on
  the test system have been allocated.}
  \label{fig:PointVsGrid}
\end{figure}

\begin{figure}
  \centering
  \includegraphics[width=0.95\textwidth]{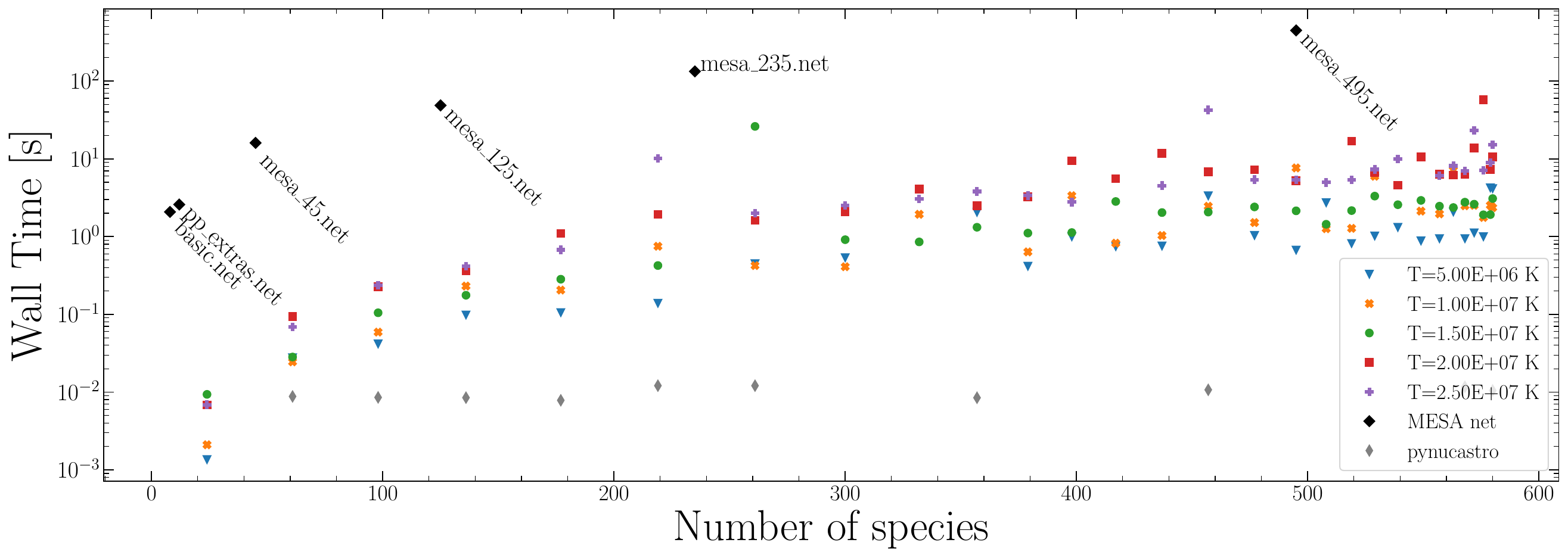}
  \caption{Wall time for a 10Gyr evaluation of \gridfire with an initial GS98
  solar composition and run at $\rho = 160$ g cm$^{-3}$. Each colored data point
  corresponds to a particular depth of network construction (i.e. the number of
  stages of graph traversal) at a particular temperature. Each black diamond
  represents the timing information for a \mesa net run while each grey narrow
  diamond corresponds to a \pynucastro run. It should be noted that the first
  set of datapoints for \gridfire (a construction depth of 1) do not include the
  full proton-proton chain. Therefore the first usable network size for main
  sequence stellar structure is the second set of datapoints, a construction
  depth of 2. The majority of tests in this paper use a construction depth of 3.}
  \label{fig:MESA_GF_Time}
\end{figure}

We benchmark \texttt{PointSolver}'s performance vs
\texttt{GridSolver}'s performance (Figure \ref{fig:PointVsGrid}). We find that,
as expected, \texttt{GridSolver} significantly reduces overall runtime. We see
that on test hardware --- a 2024 model M4 Max MacBook Pro --- the median wall
time \texttt{GridSolver} takes to evolve a network for $N$ zones is 10\%
(arithmetic mean of 12\%) that of \texttt{PointSolver}, this corresponds to a
factor of 10 speedup. The discrepancy from the full theoretical factor of 16
speedup is attributable to a number of convergent factors. First, not all cores on
the test hardware have equivalent performance, second, \gridfire is potentially
saturating the test hardware's memory bandwidth. Given those limitations a factor
of 10 speedup is likely the maximum we can expect to realize through naive
parallelism.

In addition to the number of zones other factors that influence runtime
include network size and temperature. Testing indicates that \gridfires wall
time performance scales similarly with network size when compared to \mesas
\texttt{net} module but with a constant multiplicative offset such that for the
same network size GridFire is generally two order of magnitude faster than
\mesas \texttt{net} module (Figure \ref{fig:MESA_GF_Time}). On the other hand,
\gridfire is generally much slower than \pynucastro. These discrepancies are not
surprising given the architectural differences between these three solvers. In
the case of \mesas \texttt{net} module we time this by calling the fortran
\texttt{system\_clock} subroutine before and after the call to \mesas
\texttt{net\_1\_zone\_burn} in BBQ's \texttt{lib\_hydrostatic.f90}. Further, we
exclude the first call to \texttt{net\_1\_zone\_burn} from the timing
calculation to avoid polluting timing data with file I/O as \mesa loads rate
data from its cache. Finally, in order to make the comparison as meaningful as
possible we disable eos calls inside of \texttt{net\_1\_zone\_burn} as \gridfire
does not make any similar calls. Diff files which readers may use to replicate
these runs can be found in the \gridfire repositories validation/diff directory.
Generally \mesa is conservative in regards to using a stale jacobian as it is
optimized for a highly-coupled environment. Therefore, it is not surprising that
\mesa takes longer per step for similar network sizes than \gridfire does.  On
the other hand, \pynucastro statically generates its networks and makes heavy
use of just-in-time (JIT) compilation. Therefore, each timestep with \pynucastro
collapses to a non-branching set of matrix operations; wheres, \gridfire
contains non-trivial inter-timestep logic. It should be noted that when timing
\pynucastro runs we evaluate the network twice and report the time for the
second evaluation. Due to the time required to run JIT compilation the first
network evaluation may take on the order of 10s of seconds and is not
representative of SSE use cases where a network will be evaluated many times.

\subsection{Memory Footprint}\label{sec:performance:memory}
\begin{figure}
  \centering
  \includegraphics[width=0.95\textwidth]{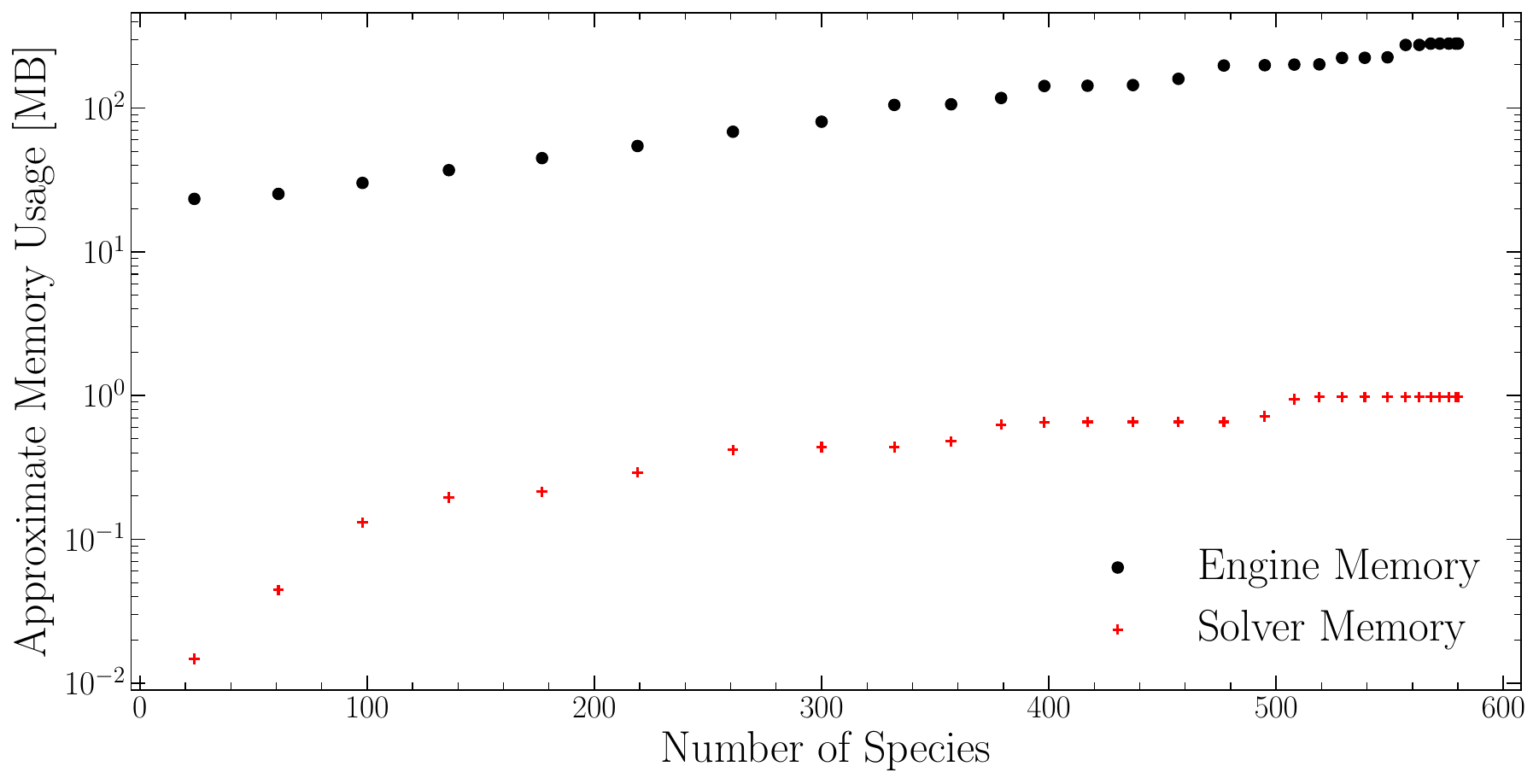}
  \caption{Approximate memory usage of \gridfire over a variety of construction
  depths. Note that this is approximate as it is non trivial to measure exact
  memory usage in C++. This measurement was taken by overloading the new and
  delete operators such that they record all allocation events and the number of bytes requested.}
  \label{fig:GF_MEM_Usage}
\end{figure}

\gridfire makes heavy use of data embedded directly into the binary. This
strategy was chosen in order to prevent common issues with locating data files
on user computers, especially when building external linkages (see
\ref{sec:usage:extension}). The major downsides to this approach are increased
binary size (a \gridfire compiled library generally sits around 25MB, though
this varies from compiler-to-compiler and system-to-system) and potentially
increased memory usage. Further, nuclear networks generally have the potential
to be memory intensive. Since we must track the relations between each network
species the size of various system matrices, such as the Jacobian matrix, scales
like the square of the number of species in the network. Finally, \gridfires use
of auto-differentiation requires in-memory storage of the computational graph
mapping each independent to each dependent variable. For all of these reasons we
do not expect \gridfire to be a memory-light library. Memory profiling shows
the expected quadratic memory use scaling with number of network species in
engine memory usage along with a much flatter memory usage profile for the
solver (Figure \ref{fig:GF_MEM_Usage}). This usage profile is in line with the
expected architecture of programs using \gridfire. That is to say that we expect
programs to hold a single engine and many solvers, therefore it is important that
the majority of memory use be on the engine side so that it is not replicated to
across solvers. The tests presented in this paper have exclusively used a network
construction depth of 3, which corresponds 136 species, 579 reactions, and an
engine memory usage of 30 MB. As said this is not a memory-light program;
however, that is fundamental to this class of problems. As a comparison tools
such BBQ using \mesas \texttt{net} module consume roughly 1.5 GB of memory when using
\texttt{mesa\_125.net}. While this is not a rigorous comparison, as \mesa
performances many additional allocations to support its much wider science goals,
it does highlight that memory usage is often high for nuclear networks.

\section{Conclusions}\label{sec:conclusions}

In this article we have presented \gridfire, a novel dynamic nuclear network
built to support extensibility and with a focus on ease of use. \gridfire can
automatically construct network topologies from a limited seed composition and
has robust support for automatic network optimization and simplification.
Further, \gridfires engine view-based architecture allows researchers to
implement domain specific features without a detailed knowledge of \gridfires
underlying code. In addition to engines, \gridfire bundles two different solvers
which can be used to integrate either single zone, or independent multi-zone
networks over gigayear timescales. We have validated \gridfires results against
well-tested nuclear networks: \texttt{pynucastro} and \mesas
\texttt{net} module, and find that \gridfire predicts kinetics which
are in good agreement with both networks given the differences in underlying
rate tables. Finally, we have demonstrated a simple BBN model which, despite
its simplicity, reproduces BBN predictions well.

Future versions of \gridfire will both expand its current capabilities ---
including adding additional screening prescriptions, robustly testing and
expanding into higher energy regimes where photodisintegration is relevant, and
relaxing the thermodynamically static approximation --- and both facilitate and
test efforts to incorporate \gridfire into existing stellar structure and
evolution programs.  Presently \gridfire will be most straightforward to
incorporate into code bases which adopt an operator splitting scheme to solve
the structure equations and microphysics. However, given \gridfires flexible
architecture we anticipate that it will also be possible to make use of
\gridfire physics engines within fully coupled structure solvers.

\section*{Acknowledgement}
\gridfire is developed as a part of the 4D-STAR collaboration
(https://4D-STAR.org). 4D-STAR is funded by European Research Council (ERC)
under the Horizon Europe programme (Synergy Grant agreement No. 101071505:
4D-STAR). Work for this project is funded by the European Union. Views and
opinions expressed are however those of the author(s) only and do not
necessarily reflect those of the European Union or the European Research
Council. We would like to thank the reviewer for their careful reading and
critique of this work. Further, we would like to thank Keighley Rockcliffe,
Rayna Rampalli, and Fen Halstead for useful discussion related to this work.
This work has made use of the NASA Astrophysical Database System (ADS). 

\appendix

\section{Usage}\label{appendix:usage}

\subsection{Policy System}
\gridfires design calls for maximum flexibility and automation, Specifically it
was a major design goal that a priori knowledge of neither network topology nor
relevant species would be required from the user. Generally, \gridfire
achieves this goal. However, there is a danger with such automation.
Specifically, that users might evolve some network assuming it represents some
physical system while in reality the inputs they provided were not sufficient to
constrain that system. For example, consider the case of an engine built from a
seed composition comprising only $^{1}$H and $^{4}$He. In order to maintain
efficiency \gridfires primary engine, \texttt{GraphEngine}, allows the user to
specify network build depth (the maximum number of layers that the breadth first
search will descend during initial network construction). This allows the user
to explicitly exclude extremely deep species which are only produced after
multiple burning stages and can dramatically increase network performance
(primarily by reducing the size of the Jacobian). Starting from a seed
composition of just two species while using a network build depth of say three, can
easily lead to certain relevant reactions being excluded. For example, in the
above case the CNO cycle would not be closed and thus the network could not
accurately model any systems where CNO was a relevant reaction pathway. 

\gridfire addresses this challenge through its policy system. Policies are
in-code directives defining some set of minimum required physics. These
objects act as the primary expected channel for users to construct engines as
they provide runtime validation of network topology. The
\texttt{MainSequenceNetworkPolicy} for example enforces that all branches of the
proton-proton chain and all branches of the CNO cycle must be present for a
network to be considered valid. If a user were to try to construct a
\texttt{MainSequenceNetworkPolicy} with a seed composition and network build
depth that resulted in some subset of either the proton-proton chain or CNO
cycles not being present then \gridfire will immediately throw an exception and
alert the user exactly which component of the policy has been violated. We
currently only bundle \texttt{MainSequenceNetworkPolicy} with \gridfire;
however, as with the rest of \gridfire, the policy system has been designed to
be easily extended.

\subsection{Extension}\label{sec:usage:extension}
\gridfire has been, from the ground up, designed to be user extensible from
either C++ or Python. All objects in \gridfire are implemented as derived
classes from pure virtual, interface, classes. For example, a user may derive the
\texttt{DynamicEngineViewWrapper} pure virtual class from \gridfire in order to
implement new physics. As a contrived example consider the case where this new
physics is as simple as saying that the Jacobian column for $^{12}$C is all 0.
Implementing this would be as simple as overriding the
\texttt{generateJacobianMatrix} methods such that they call the base engine and 
then zero out the $^{12}$C column prior to returning the Jacobian to the caller.

A more realistic example might build off of the spectral solver discussed in \S 
\ref{sec:solvers:spectral}. Wheres the example spectral solver implicitly took
an operator-splitting approach to SSE (in so far as
that solver has no ability to couple structural calculations with network
evolution), a solver may be implemented which is coupled. Alternatively, an
equation of state may be incorporated into some new solver to relax the
thermodynamically static assumption \gridfire currently makes.  So long as these
expansions implement the same minimum interface as \gridfire defines in its
abstract base classes they can be used with all other \gridfire methods.

Finally, while extension through C++ code will demonstrate by far the most
performance, every abstract base class has had a so called ``trampoline class''
implemented in the Python bindings. This allows for Python classes to override
abstract C++ classes and be passed in place of C++ objects from Python code. This
does have the disadvantage of a substantial performance overhead (Python 
is notoriously slow); however, for cases where most of the heavy computational work
can still be offloaded to C++, this provides an avenue for non C++ developers
to make substantial modifications to \gridfire such that it might fit their
specific needs.

\subsection{External Use}
\gridfire is entirely written in C++ and as such inherits the famously high
performance of that language. However, C++ is not a widely used language in
astronomy, certainly not outside of computational astronomy software
development.  Other languages, Python foremost among them and --- to a lesser
extent --- Fortran, dominate software development for astrophysics. While
\gridfire may be useful for the 4D-STAR collaboration as a purely C++ based
software stack, we aim to make this library broadly useful to the community. As
such significant effort has been put into developing bridges to other
languages. Currently we have demonstrated full \gridfire capabilities in Python
--- including use and extension and limited \gridfire abilities --- use but not
extension --- in C, Fortran, and Javascript. We anticipate that \gridfire will
be primarily used as a Python module. All external language bindings
can be found in the primary GridFire repository.

\subsubsection{Python Interface}\label{sec:usage:Python}
Python is a widely language in astronomy and astrophysics today. As
such Python is the only language, other than C++, which will receive full
\gridfire bindings. That is to say that every single class, function, enum,
exception, and variable available in the \gridfire C++ API is made available in
the Python API. This is done using \texttt{pybind11} \citep{pybind11}. Python is
generally expected to be the language which users interact with \gridfire
through. In addition to bindings, precompiled Python wheels have been built for
a number of systems and registered on the Python package index (pypi) such that
familiar command \texttt{pip install gridfire} will work for any macos ($\geq$
15.0) or Linux system (more recent than Ubuntu 18.04) without requiring time-
and resource-consuming compilation.

\subsubsection{C API}\label{sec:usage:capi}
For historical reasons C++ does not provide a stable application binary interface (ABI);
rather, different implementations of C++ standards-compliant compilers and
standard template libraries implement their own ABIs. This poses a problem when
interfacing to other languages. In order to resolve this we bind a
subset of the \gridfire API to a stable C API. Note that this subset is
sufficient for use but does not allow the caller to hook into \gridfires
extension capabilities (such as defining new \texttt{EngineView}s). Overall, the
C API presents seven functions (Table \ref{tab:CAPI}) which encapsulate all
setup, work, and tear down functionality of \gridfire.

\begin{table}[htb!]
\centering
\small
\begin{tabular}{p{0.43\textwidth} p{0.52\textwidth}}
\hline
\textbf{Method Name} & \textbf{Purpose} \\
\hline
\texttt{gf\_init} & Initialize the context pointer (\texttt{ctx}) which stores all \gridfire state information. \\
\texttt{gf\_get\_last\_error\_message} & Access a human-readable representation of the last error \gridfire threw.  \\
\texttt{gf\_register\_species} & Tell \gridfire which species the user will be providing molar abundance values for. \\
\texttt{gf\_construct\_engine\_from\_policy} & Build the policy and engine stack. \\
\texttt{gf\_construct\_solver\_from\_engine} & Build a single- or multi-zone solver from the constructed engine stack. \\
\texttt{gf\_evolve} & Integrate a given abundance through time with the constructed solver. \\
\texttt{gf\_free} & Clean up all resources used by \gridfire to ensure no memory leaks. \\
\hline
\end{tabular}
\caption{Major components implemented by \gridfire. Note that these are not exhaustive but do represent the most important subset of abstract types.}
\label{tab:CAPI}
\end{table}

Any C code including the \texttt{gridfire/extern/gridfire\_extern.h} header and
linking against the \texttt{gridfire\_mod} shared library will be able to access
these functions. In addition to bringing C support, the presence of a C API
allows \gridfire to be called reliably from any languages with C
interoperability such as Zig\footnote{https://ziglang.org/} or Rust\footnote{https://rust-lang.org/} through a foreign function interface. Example of
C API usage may be found in \gridfires documentation.

\subsubsection{Fortran Interface}\label{sec:usage:Fortran}
Many high performance astrophysics code bases are written in Fortran \citep[e.g
\mesa,][]{Paxton2011} and as such it is useful to expose a \gridfire interface
which can be used from Fortran. We leverage the C API described in
\ref{sec:usage:capi} and Fortran 2003's \texttt{iso\_c\_bindings} to bind a
\texttt{GridFire} type in Fortran. As this is effectively a wrapper around the C
API usage will look very similar. Examples may be found in the \gridfire
documentation. 

\subsubsection{Javascript and Web Assembly}
\gridfire can compile to a web assembly target (WASM) which opens the door for
web-native execution of nuclear-burning simulations. This is likely not of
direct interest for research applications. However, students
using simulation tools often face challenges with software installation. We
have used the compiler toolchain Emscripten to build a set of Javascript
bindings equivalent the the C API described in Table \ref{tab:CAPI} which allow
for \gridfire to be incorporated into a website easily\footnote{An example of
what a \gridfire website might look like is presented at
https://gridfire.algebrist.com}. By packaging \gridfire into a static web-site we
open the door to classroom exploration and use without any expectation of
students installing anything more uncommon than a modern web browser.

\section{\gridfire Control Flow}
Generally astrophysical simulation software is complex. Here we provide a
set of simplified flow charts meant to clarify the structure of how \gridfire
works. 
\subsection{Prototypical Internal Control Flow}
\begin{figure}[htb!]
  \resizebox{\linewidth}{!}{
    \begin{tikzpicture}[
    node distance=1.5cm and 1cm,
    font=\sffamily\footnotesize,
    >=Latex,
    startstop/.style={rectangle, rounded corners, minimum width=2.5cm, minimum height=0.6cm, inner sep=4pt, text centered, draw=black, fill=red!10},
    process/.style={rectangle, minimum width=3.2cm, minimum height=0.6cm, inner sep=4pt, text centered, draw=black, fill=blue!10},
    decision/.style={diamond, aspect=2.5, minimum width=2.5cm, minimum height=0.6cm, inner sep=2pt, text centered, draw=black, fill=green!10},
    io/.style={trapezium, trapezium left angle=70, trapezium right angle=110, minimum width=2.8cm, minimum height=0.6cm, inner sep=4pt, text centered, draw=black, fill=orange!10},
    stack/.style={rectangle, minimum width=3cm, minimum height=0.6cm, inner sep=3pt, text centered, draw=black, fill=purple!10, dashed},
    line/.style={draw, ->, thick},
    label_text/.style={font=\scriptsize\itshape, text=gray}
]

\node (user_start) [startstop] {User Start};
\node (def_policy) [process, below of=user_start] {Define \textbf{NetworkPolicy}\\(Pass Seed Composition)};
\node (construct) [process, below of=def_policy] {Call \texttt{policy.construct()}};

\node (policy_build) [process, below of=construct, text width=3.5cm] {\textbf{Policy Internal Logic}:\\Validate \& Build Engine Stack\\Returns: \texttt{EngineView} (Top)};

\node (solver_init) [process, below of=policy_build, yshift=-0.2cm] {Instantiate \textbf{Solver}\\(Single/Multi Zone)};
\node (build_netin) [io, below of=solver_init] {Build \texttt{NetIn}\\(T, $\rho$, time, comp)};

\node (eval) [process, below of=build_netin] {Call \texttt{solver.evaluate(NetIn)}};

\node (trig_build) [process, below of=eval] {Build Trigger \& Init Resources};

\node (update_call) [process, below of=trig_build, fill=yellow!20, thick] {\textbf{Call project()}\\Sync Engine Stack};

\node (timestep) [process, below of=update_call] {Start Time-Step};
\node (calc_rhs) [process, below of=timestep] {Evaluate $\frac{dY_{i}}{dt}$, $\epsilon_{nuc}$, $\epsilon_{\nu}$, \& $J$};
\node (check_trig) [decision, below of=calc_rhs, text width=1.8cm] {Check Trigger};
\node (check_term) [decision, below of=check_trig, yshift=-0.3cm, text width=2cm] {Max Time / Error?};

\node (marshal) [process, below of=check_term, yshift=-0.3cm] {Marshal Output (\texttt{NetOut})};
\node (end) [startstop, below of=marshal] {Return to User};

\node (stack_top) [stack, right=2.5cm of update_call, fill=purple!20, draw] {Top: \textbf{AEV}};
\node (stack_mid) [stack, below=1cm of stack_top] {Mid: \textbf{MSPEV}};
\node (stack_base) [stack, below=1cm of stack_mid] {Base: \textbf{GraphEngine}};

\node (base_logic) [process, right=0.75cm of stack_base, text width=2.8cm, font=\scriptsize, inner sep=2pt] {Base Update:\\Ensure Species Exist\\Init ScratchPad};
\node (mid_logic) [process, right=0.75cm of stack_mid, text width=2.8cm, font=\scriptsize, inner sep=2pt] {Partitioning (QSE):\\Update Comp\\Update ScratchPad};
\node (top_logic) [process, right=0.75cm of stack_top, text width=2.8cm, font=\scriptsize, inner sep=2pt] {Pruning:\\Cull pathways\\Finalize Topology};

\begin{scope}[on background layer]
    \node (stack_box) [draw=gray, dotted, fit=(stack_top) (stack_base) (base_logic) (top_logic), inner sep=0.75cm, rounded corners, label={[anchor=north west, font=\scriptsize]north west:\textbf{Engine Stack Recursion}}] {};
\end{scope}

\draw [line] (user_start) -- (def_policy);
\draw [line] (def_policy) -- (construct);
\draw [line] (construct) -- (policy_build);
\draw [line] (policy_build) -- node[right, label_text] {Returns Engine Ref} (solver_init);
\draw [line] (solver_init) -- (build_netin);
\draw [line] (build_netin) -- (eval);
\draw [line] (eval) -- (trig_build);
\draw [line] (trig_build) -- (update_call);
\draw [line] (top_logic) -- (stack_top);

\draw [->, thick, >=Latex] (update_call.south east) -- node[below, label_text, align=center, font=\tiny] {Solver calls Update\\of top most view} (stack_top.south west);
\draw [<-, thick, >=Latex] (update_call.north east) -- node[above, label_text, align=center, font=\tiny] {Update returns scratch \\\& composition} (stack_top.north west);

\draw [line] (stack_top) -- (stack_mid);
\draw [line] (stack_mid) -- (stack_base);

\draw [line] (stack_base) -- (base_logic);

\draw [line] (base_logic.north) -- (mid_logic.south);
\draw [line] (mid_logic.north) -- (top_logic.south);

\draw [line] (update_call) -- (timestep);
\draw [line] (timestep) -- (calc_rhs);
\draw [line] (calc_rhs) -- (check_trig);

\draw [line] (check_trig.west) -- node[above, font=\scriptsize, xshift=0.2cm] {Trigger Fired} ++(-1.5cm,0) |- (update_call.west);
\draw [line] (check_trig.south) -- node[right, font=\scriptsize] {No Change} (check_term.north);

\draw [line] (check_term.west) -- node[above, font=\scriptsize, xshift=0.2cm] {Continue} ++(-2.2cm,0) |- (timestep.west);
\draw [line] (check_term.south) -- node[right, font=\scriptsize] {Finished} (marshal.north);
\draw [line] (marshal) -- (end);

\matrix [draw, anchor=north east, fill=white, inner sep=5pt, nodes={font=\scriptsize, anchor=west}, row sep=2pt] at ([yshift=8cm]stack_box.north east) {
    \node[shape=rectangle, fill=red!10, draw, minimum width=1em, minimum height=1em] {}; & \node{Start/End State}; \\
    \node[shape=rectangle, fill=blue!10, draw, minimum width=1em, minimum height=1em] {}; & \node{Process/Action}; \\
    \node[shape=diamond, aspect=2, fill=green!10, draw, minimum width=1em, scale=0.5] {}; & \node{Decision Point}; \\
    \node[shape=trapezium, fill=orange!10, draw, minimum width=1em, scale=0.7] {}; & \node{Input/Output}; \\
    \node[shape=rectangle, fill=purple!10, dashed, draw, minimum width=1em, minimum height=1em] {}; & \node{Engine Element}; \\
    \node[shape=rectangle, fill=yellow!20, draw, minimum width=1em, minimum height=1em] {}; & \node{Key Sync Point}; \\
    \node{}; & \node{\textbf{Acronyms}}; \\
    \node{}; & \node{AEV: AdaptiveEngineView}; \\
    \node{}; & \node{MSPEV: MultiscalePartitioningEngineView}; \\
    \node{}; & \node{QSE: Quasi-Steady State Equilibrium}; \\
};

\end{tikzpicture}
  }
  \caption{Sketch of a prototypical \gridfire run for a single zone solver.}
  \label{fig:gridfire_flow_chart}
\end{figure}
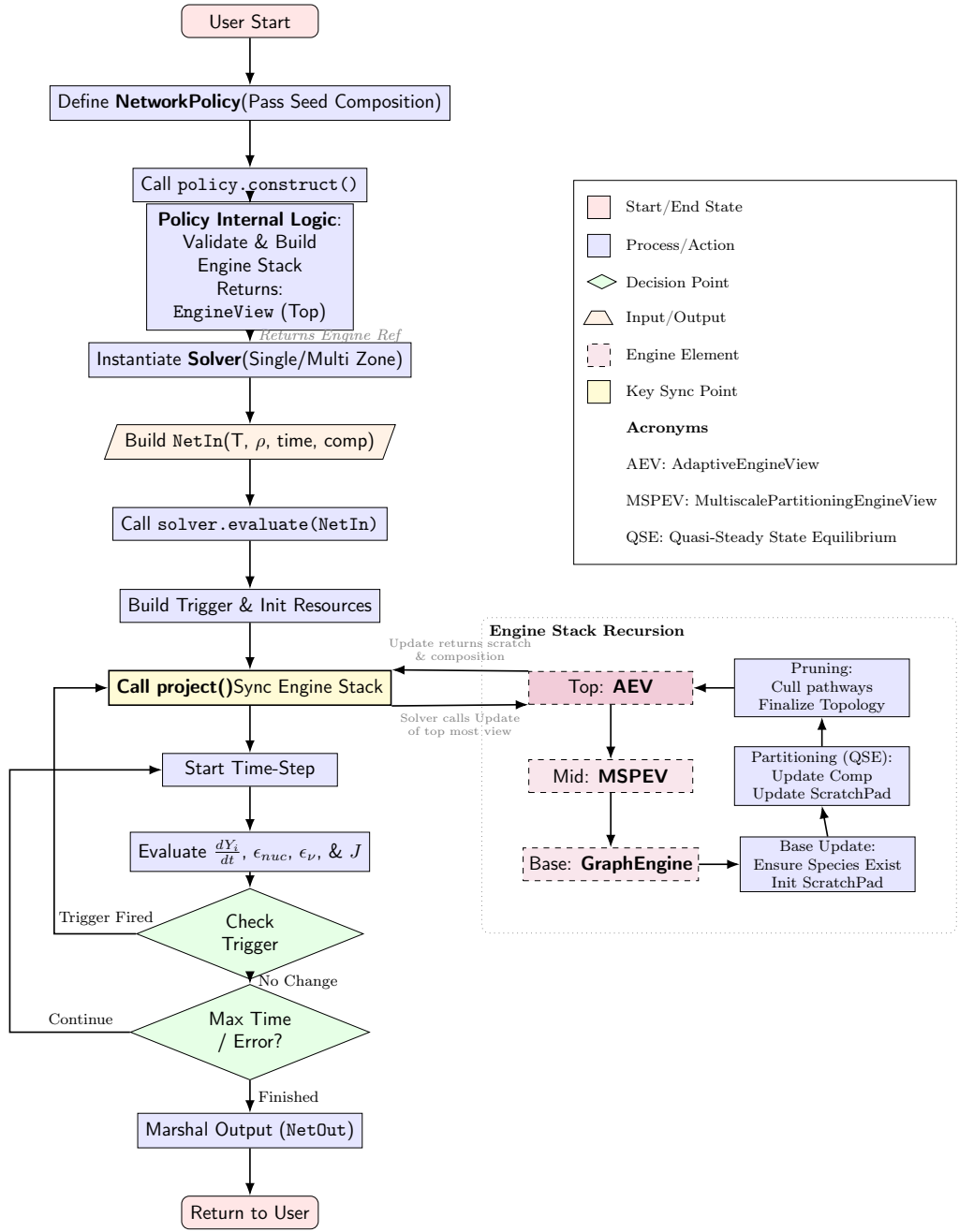

Figure \ref{fig:gridfire_flow_chart} provides a prototypical examples of how \gridfire runs. This figure is 
focused on developing an understanding of \gridfires internal workings rather than the process of using \gridfire.
\subsection{Engine View Projection Control Flow}
Here we present diagrams of the control flow for each engine view's project() method (Figure \ref{fig:qseev_control_flow}).

\begin{figure}
  \centering
  \resizebox{0.8\linewidth}{!}{\begin{tikzpicture}[
    node distance=0.8cm and 1.0cm,
    font=\sffamily\footnotesize,
    >=Latex,
    startstop/.style={
        rectangle, rounded corners=3mm, minimum width=3cm, minimum height=0.8cm,
        text centered, draw=black!80, fill=red!10, thick
    },
    process/.style={
        rectangle, minimum width=3.5cm, minimum height=0.8cm,
        text centered, draw=black!80, fill=blue!5, thick
    },
    subroutine/.style={
        rectangle, minimum width=3.5cm, minimum height=0.8cm,
        inner xsep=4mm, 
        text centered, draw=black!80, fill=blue!15, thick,
        path picture={
            \draw[black] ([xshift=2mm]path picture bounding box.north west) -- ([xshift=2mm]path picture bounding box.south west);
            \draw[black] ([xshift=-2mm]path picture bounding box.north east) -- ([xshift=-2mm]path picture bounding box.south east);
        }
    },
    decision/.style={
        diamond, 
        aspect=1.8, 
        minimum width=1.5cm, 
        minimum height=0.8cm,
        text centered, draw=black!80, fill=green!10, inner sep=1pt
    },
    io/.style={
        trapezium, 
        trapezium left angle=80, trapezium right angle=100, 
        minimum width=0cm, 
        minimum height=0.8cm,
        text centered, draw=black!80, fill=orange!10,
        inner xsep=4mm 
    },
    container/.style={
        draw=gray!50, dashed, rounded corners, inner sep=0.5cm, fill=white, fill opacity=0.3
    },
    line/.style={draw, ->, thick, black!80},
    connector/.style={draw, ->, thick, black!80, rounded corners},
    label_text/.style={font=\scriptsize\itshape, text=gray!80, anchor=west}
]

    \node (start) [startstop] {\textbf{Start}: \texttt{project()}};
    \node (base_proj) [process, below=0.6cm of start] {Base Engine Projection};
    
    \node (prime) [subroutine, below=1.0cm of base_proj] {Prime Engine \& Clear State};
    \node (partition_ts) [subroutine, below=0.6cm of prime] {Partition by Timescale};
    
    \node (id_slowest) [process, below=0.8cm of partition_ts, xshift=-2.5cm] {Identify Slowest Pool};
    \node (set_dynamic) [io, below=0.6cm of id_slowest] {Set Core Dynamic Species};

    \node (connectivity) [subroutine, below=0.8cm of partition_ts, xshift=4.5cm] {Pool Connectivity Analysis};
    \node (construct) [process, below=0.6cm of connectivity] {Construct Candidate Groups};
    
    \node (validate_1) [decision, below=0.8cm of construct] {Flux Validated?};
    \node (prune) [process, below=0.8cm of validate_1] {Prune via Log-Flux};
    \node (validate_2) [decision, below=0.8cm of prune] {Re-Validated?};
    \node (merge) [process, below=0.8cm of validate_2] {Merge Coupled Groups};

    \path let \p1=(set_dynamic), \p2=(merge) in coordinate (bottom_center) at ({(\x1+\x2)/2}, \y2);
    
    \node (finalize) [process, below=1.5cm of bottom_center] {Finalize Species Sets};
    \node (create_solvers) [process, below=0.6cm of finalize] {Create QSE Solvers};
    \node (solve_eq) [subroutine, below=0.6cm of create_solvers] {Solve QSE Abundances};
    
    \node (clear_cache) [process, below=0.6cm of solve_eq] {Clear Composition Cache};
    \node (end) [startstop, below=0.8cm of clear_cache] {\textbf{Return}: Composition};

    \draw [line] (start) -- (base_proj);
    \draw [line] (base_proj) -- node[right, font=\scriptsize, text=gray] {Call partitionNetwork()} (prime);
    \draw [line] (prime) -- (partition_ts);
    
    \draw [connector] (partition_ts.south) -- +(0,-0.3) -| (id_slowest.north);
    \draw [connector] (partition_ts.south) -- +(0,-0.3) -| (connectivity.north);

    \draw [line] (id_slowest) -- (set_dynamic);
    
    \draw [line] (connectivity) -- (construct);
    \draw [line] (construct) -- (validate_1);
    
    \draw [line] (validate_1) -- node[right, font=\scriptsize] {YES} (prune);
    \node (discard_1) [right=0.4cm of validate_1, font=\scriptsize, text=gray, align=left] {Discard\\Group};
    \draw [connector] (validate_1.east) -- (discard_1.west);

    \draw [line] (prune) -- (validate_2);
    
    \draw [line] (validate_2) -- node[right, font=\scriptsize] {YES} (merge);
    \node (discard_2) [right=0.4cm of validate_2, font=\scriptsize, text=gray, align=left] {Discard\\Group};
    \draw [connector] (validate_2.east) -- (discard_2.west);

    \draw [connector] (set_dynamic.south) |- (finalize.west);
    \draw [connector] (merge.south) |- (finalize.east);

    \draw [line] (finalize) -- (create_solvers);
    \draw [line] (create_solvers) -- (solve_eq);
    \draw [line] (solve_eq) -- (clear_cache);
    \draw [line] (clear_cache) -- (end);

    \begin{scope}[on background layer]
        \node (pipeline_group) [container, fit=(connectivity) (construct) (validate_1) (prune) (validate_2) (merge) (discard_1) (discard_2), 
        label={[anchor=south, rotate=90, font=\bfseries\scriptsize, text=gray, yshift=2mm]west:QSE Group Refinement}] {};
    \end{scope}

    \node [left=0.1cm of prime, font=\scriptsize, text=gray, align=right] {Step 0};
    \node [left=0.1cm of partition_ts, font=\scriptsize, text=gray, align=right] {Step 1};
    \node [left=0.1cm of id_slowest, font=\scriptsize, text=gray, align=right] {Step 2};
    \node [left=0.1cm of set_dynamic, font=\scriptsize, text=gray, align=right] {Step 3};
    \node [right=0.1cm of connectivity, font=\scriptsize, text=gray, align=left] {Step 4};
    \node [right=0.1cm of construct, font=\scriptsize, text=gray, align=left] {Step 5};
    \node [above left=0.1cm of finalize, font=\scriptsize, text=gray, align=right] {Step 6};
    \node [left=0.1cm of create_solvers, font=\scriptsize, text=gray, align=right] {Step 7};
    \node [left=0.1cm of solve_eq, font=\scriptsize, text=gray, align=right] {Step 8};

\end{tikzpicture}}
  \caption{Control flow for \qsev projection.}
  \label{fig:qseev_control_flow}
\end{figure}


\section{License}
\gridfire is licensed under the GNU General Public License v3.0. A copy of this
license is bundled with the \gridfire source code; you may also retrieve a copy
from https://www.gnu.org/licenses/gpl-3.0.en.html


\bibliography{ms}{}
\bibliographystyle{elsarticle-harv}

\end{document}